\documentclass[preprint]{aastex61}
\usepackage{longtable}
\received{}
\revised{\today}
\submitjournal{ApJ}

\shorttitle{Signatures of Diffusion in the Open Cluster M67}
\shortauthors{Souto et al.}

\begin{document}

\title{Chemical Abundances of Main-Sequence, Turn-off, Subgiant and red giant Stars from APOGEE spectra I: Signatures of Diffusion in the Open Cluster M67}

\correspondingauthor{Diogo Souto}
\email{souto@on.br, diogodusouto@gmail.com}

\author[0000-0002-7883-5425]{Diogo Souto}
\affiliation{Observat\'orio Nacional, Rua General Jos\'e Cristino, 77, 20921-400 S\~ao Crist\'ov\~ao, Rio de Janeiro, RJ, Brazil}

\author{Katia Cunha}
\affiliation{Steward Observatory, University of Arizona, 933 North Cherry Avenue, Tucson, AZ 85721-0065, USA}
\affiliation{Observat\'orio Nacional, Rua General Jos\'e Cristino, 77, 20921-400 S\~ao Crist\'ov\~ao, Rio de Janeiro, RJ, Brazil}

\author{Verne V. Smith}
\affiliation{National Optical Astronomy Observatory, 950 North Cherry Avenue, Tucson, AZ 85719, USA}

\author{C. Allende Prieto}
\affiliation{Instituto de Astrof\'isica de Canarias, E-38205 La Laguna, Tenerife, Spain}
\affiliation{Departamento de Astrof\'isica, Universidad de La Laguna, E-38206 La Laguna, Tenerife, Spain}

\author{D. A. Garc\'ia-Hern\'andez}
\affiliation{Instituto de Astrof\'isica de Canarias, E-38205 La Laguna, Tenerife, Spain}
\affiliation{Departamento de Astrof\'isica, Universidad de La Laguna, E-38206 La Laguna, Tenerife, Spain}

\author{Marc Pinsonneault}
\affiliation{Department of Astronomy, The Ohio State University, Columbus, OH 43210, USA}

\author{Parker Holzer}
\affiliation{Department of Physics and Astronomy, The University of Utah, Salt Lake City, UT 84112, USA}

\author[0000-0002-0740-8346]{Peter Frinchaboy}
\affiliation{Department of Physics \& Astronomy, Texas Christian University, Fort Worth, TX, 76129, USA}

\author{Jon Holtzman}
\affiliation{New Mexico State University, Las Cruces, NM 88003, USA}

\author{J. A. Johnson}
\affiliation{Department of Astronomy, The Ohio State University, Columbus, OH 43210, USA}

\author[0000-0002-4912-8609]{Henrik J\"onsson}
\affiliation{Lund Observatory, Department of Astronomy and Theoretical Physics, Lund University, Box 43, SE-221 00 Lund, Sweden}

\author{Steven R. Majewski}
\affiliation{Department of Astronomy, University of Virginia, Charlottesville, VA 22904-4325, USA}

\author{Matthew Shetrone}
\affiliation{University of Texas at Austin, McDonald Observatory, USA}

\author{Jennifer Sobeck}
\affiliation{Department of Astronomy, University of Virginia, Charlottesville, VA 22904-4325, USA}

\author{Guy Stringfellow}
\affiliation{Center for Astrophysics and Space Astronomy, Department of Astrophysical and Planetary Sciences, University of Colorado, Boulder, CO, 80309, USA}

\author{Johanna Teske}
\affiliation{Department of Terrestrial Magnetism, Carnegie Institution for Science, Washington, DC 20015}

\author{Olga Zamora}
\affiliation{Instituto de Astrof\'isica de Canarias, E-38205 La Laguna, Tenerife, Spain}
\affiliation{Departamento de Astrof\'isica, Universidad de La Laguna, E-38206 La Laguna, Tenerife, Spain}

\author{Gail Zasowski}
\affiliation{Department of Physics and Astronomy, The University of Utah, Salt Lake City, UT 84112, USA}

\author{Ricardo Carrera}
\affiliation{INAF - Osservatorio Astronomico di Padova, Padova, Italy}

\author{Keivan Stassun}
\affiliation{Department of Physics and Astronomy, Vanderbilt University, VU Station 1807, Nashville, TN 37235, USA}

\author{J. G. Fernandez-Trincado}
\affiliation{Departamento de Astronom\'ia, Universidad de Concepci\'on, Casilla 160-C, Concepci\'on, Chile}
\affiliation{Institut Utinam, CNRS UMR6213, Univ. Bourgogne Franche-Comt\'e, OSU THETA, Observatoire de Besan\c con, BP 1615, 25010 Besan\c con Cedex}

\author{Sandro Villanova}
\affiliation{Departamento de Astronom\'ia, Universidad de Concepci\'on, Casilla 160-C, Concepci\'on, Chile}

\author{Dante Minniti}
\affiliation{Instituto de Astrof\'isica, Pontificia Universidad Cat\'olica de Chile, Av. Vicuna Mackenna 4860, 782-0436 Macul, Santiago, Chile}

\author{Felipe Santana}
\affiliation{Universidad de Chile, Av. Libertador Bernardo O’Higgins 1058, Santiago de Chile}



\begin{abstract}
Detailed chemical abundance distributions for fourteen elements are derived for eight high-probability stellar members of the solar metallicity old open cluster M67 with an age of $\sim$4 Gyr.
The eight stars consist of four pairs, with each pair occupying a distinct phase of stellar evolution: two G-dwarfs, two turnoff stars, two G-subgiants, and two red clump K-giants.  
The abundance analysis uses near-IR high-resolution spectra ($\lambda$1.5 -- 1.7$\mu$m) from the APOGEE survey and derives abundances for C, N, O, Na, Mg, Al, Si, K, Ca, Ti, V, Cr, Mn, and Fe. 
Our derived stellar parameters and metallicity for 2M08510076+113115 suggest that this star is a solar-twin, exhibiting abundance differences relative to the Sun of $\leq$ 0.04 dex for all elements.
Chemical homogeneity is found within each class of stars ($\sim$0.02 dex), while significant abundance variations ($\sim$0.05 -- 0.20 dex) are found across the different evolutionary phases; the turnoff stars typically have the lowest abundances, while the red clump tend to have the largest. 
Non-LTE corrections to the LTE-derived abundances are unlikely to explain the differences.  
A detailed comparison of the derived Fe, Mg, Si, and Ca abundances with recently published surface abundances from stellar models that include chemical diffusion, provides a good match between the observed and predicted abundances as a function of stellar mass. 
Such agreement would indicate the detection of chemical diffusion processes in the stellar members of M67.
\end{abstract}

\keywords{infrared: star -- general: open clusters and associations stars -- stars: abundances -- Physical data and processes: diffusion}


\section{Introduction}

Messier 67 (M67, or NGC 2682), is one of the most studied Galactic open clusters, due in part to it having an age and metallicity 
that are similar to those of the Sun, thus making it a useful cluster in which to study the properties of solar twins, as well as the evolution of solar-like stars. 
Photometric studies have provided a well-determined reddening  (E(B-V)=0.041; \citealt{Taylor2007}, \citealt{Sarajedini2009}), distance modulus ($\mu$ = 9.56--9.72, or d=800--860 pc; \citealt{Yadav2008}), and age ($\sim$ 4.0 Gyrs; \citealt{Salaris2004}, \citealt{Yadav2008}, \citealt{Sarajedini2009}) for M67.

High-resolution spectroscopic abundance studies have found M67 to have a near-solar chemical composition ([Fe/H] $\sim$ 0.00; \citealt{Cohen1980}; \citealt{FoyProust1981}; \citealt{FrielBoesgaard1992}; \citealt{Tautvaisiene2000}; \citealt{Pancino2010}; \citealt{Jacobson2011}; \citealt{Onehag2014}; \citealt{Liu2016}, \citealt{Casamiquela2017}).
M67 also falls in one of the Kepler K2 Campaign fields and thus has precision photometry available for a number of its members; these data have been used in recent asteroseismology studies of the red giants in M67 (e.g., \citealt{Gonzalez2016}; \citealt{Stello2016}; \citealt{Leiner2016}). 

In addition to containing stars that are similar to the Sun in mass, age, and chemical composition, M67 provides a laboratory in which to explore the properties of stars over a range of evolutionary phases: from dwarfs on the main sequence, up to the turnoff, through the subgiant branch, and onto the red giant branch (with these stars having very nearly the same ages and initial chemical compositions).  Quantitative high-resolution spectroscopic analyses can be used to reveal if there are chemical abundance differences between the M67 members in the different evolutionary phases, as well as determining whether these differences are real, or if they reveal systematic differences induced by the analysis techniques themselves. Acccurate chemical abundances of a large number of elements can be used to test models of chemical diffusion in stars (e.g., \citealt{Michaud2015}).
The recent studies by \cite{Onehag2014}, \cite{Blanco-Cuaresma2015}, and \cite{BertelliMotta2017} find chemical inhomogeneities for some elements that could be an indication that diffusion mechanisms may be at work and may be detectable in M67 stars.

Chemical abundance variations that result from diffusion have been probed in the metal-poor globular cluster NGC 6397 ([Fe/H]= -2.1), in a series of papers by \cite{Korn2007}, \cite{Lind2008}, and \cite{Nordlander2012}.
These papers studied cluster stars at the turnoff point (TOP), on the subgiant branch (SGB), at the base of the red giant branch (bRGB), and on the red giant branch (RGB) using chemical abundances of Li, Mg, Ca, Ti, Cr, and Fe.  
Chemical abundances were found to be dependent on the evolutionary stage and these were compared to diffusion models, with overall approximate agreement between observed and model abundances, particularly for Mg and Fe.

Diffusion model predictions specific to the solar-metallicity and age regime of M67 have been presented by \cite{Michaud2004}, as well as most recently by \cite{Dotter2017}.  
In both studies, diffusion effects near the turnoff in M67 have been found to be as large as $\sim$0.1 dex when compared to cooler main sequence stars or stars on the RGB (where convection erases the abundance patterns created by diffusion), especially for certain elements such as Fe or Mg. 
Specific patterns in the abundance distributions of certain elements, such as Ca and Fe are also predicted to exhibit detectable variations, suggesting that M67 is a key open cluster in which to probe and test models of diffusion processes.

In this study, a small sample of M67 members that were observed as part of the Apache Point Observatory Galactic Evolution Experiment (APOGEE; \citealt{Majewski2017}) survey are analyzed using its high-resolution, near-infrared (NIR) spectra to derive accurate chemical abundances for a large number of elements.  
The studied sample spans a range of evolutionary phases consisting of G-dwarfs, G-turnoff stars, G-subgiants, and K-giants. 
The detailed chemical abundances derived for these stars are used to investigate possible abundance inhomogeneities in M67 as a function of stellar class and to determine whether such abundance variations can be explained by a physical process, such as diffusion in the stellar envelope, or reflect systematic effects associated with the analysis techniques.
The recent studies of \cite{Bovy2016} and \cite{Price-Jones2018} using APOGEE spectra found the tightest constraints on the chemical homogeneity of M67 red giants. This evidence of chemical homogeneity in red giants is an important starting point to investigate possible diffusion effects and assign observed abundance differences to the effects of stellar evolution.

Section 2 discusses the details of the APOGEE survey and spectra, while Section 3 presents the determination of stellar parameters (effective temperature, $T_{\rm eff}$, and surface gravity, log $g$), along with the chemical abundances.  
The chemical abundance distributions and possible variations are discussed in Sections 4 and 5, with the summary of results in Section 6.

\section{The APOGEE Spectra and the Sample}

The spectra analyzed in this work are from the SDSS-IV/APOGEE2 survey (\citealt{Blanton2017}, \citealt{Majewski2017}). The APOGEE instrument is a cryogenic multi-fiber spectrograph (300-fibers; \citealt{Wilson2010}) on the SDSS 2.5-m telescope at the Apache Point Observatory \citep{Gunn2006}. 
The survey observations consist of high-resolution (R=$\lambda$/$\Delta$$\lambda$ $\sim$ 22,500) spectra of stars, primarily red giants, but also of stars in other evolutionary stages (see \citealt{Zasowski2013}), in the near infrared ($\sim$ $\lambda$1.50 $\mu$m - $\lambda$1.70 $\mu$m) with the ultimate goal of exploring the chemical evolution of the stellar populations in the Milky Way.

Our sample contains eight stars that are deemed to be members of the M67 open cluster. 
The APOGEE spectra of the sample stars were reduced automatically by the APOGEE pipeline (\citealt{Nidever2015}, \citealt{Holtzman2015}) and then analyzed manually  to extract detailed chemical abundances.
We selected targets strategically to sample a range in effective temperatures and surface gravities that are representative of stars on the main sequence and in more advanced phases of evolution: two G-dwarfs, two G-type turnoff stars, two G-type subgiants, and two K-type red giants. All targets are in the proper motion study of \cite{Yadav2008} and have probabilities of membership higher than 91\% (Table 1). Their measured radial velocities from the APOGEE spectra are also presented in Table 1, and these are consistent with probable cluster membership \citep{Geller2015}.
Figure 1 shows the color-magnitude diagram for the near-infrared (NIR) 2MASS data \citep{2MASS}, with $H_{0}$ plotted versus ($J$-$K_{s}$)$_{0}$. The eight target stars are shown as filled red symbols.
We also show, as grey dots, the 536 stars from the M67 field observed by APOGEE (which include additional M67 members) and two isochrones for an age of 4 Gyr, (m-M)$_{0}$ = 9.60, and [Fe/H] = 0.00 from PARSEC (\citealt{Bressan2012}), and from MIST (\citealt{Dotter2016}, \citealt{Choi2016}).

\begin{figure*}
\figurenum{1}
\begin{center}
\includegraphics[width=0.7\linewidth]{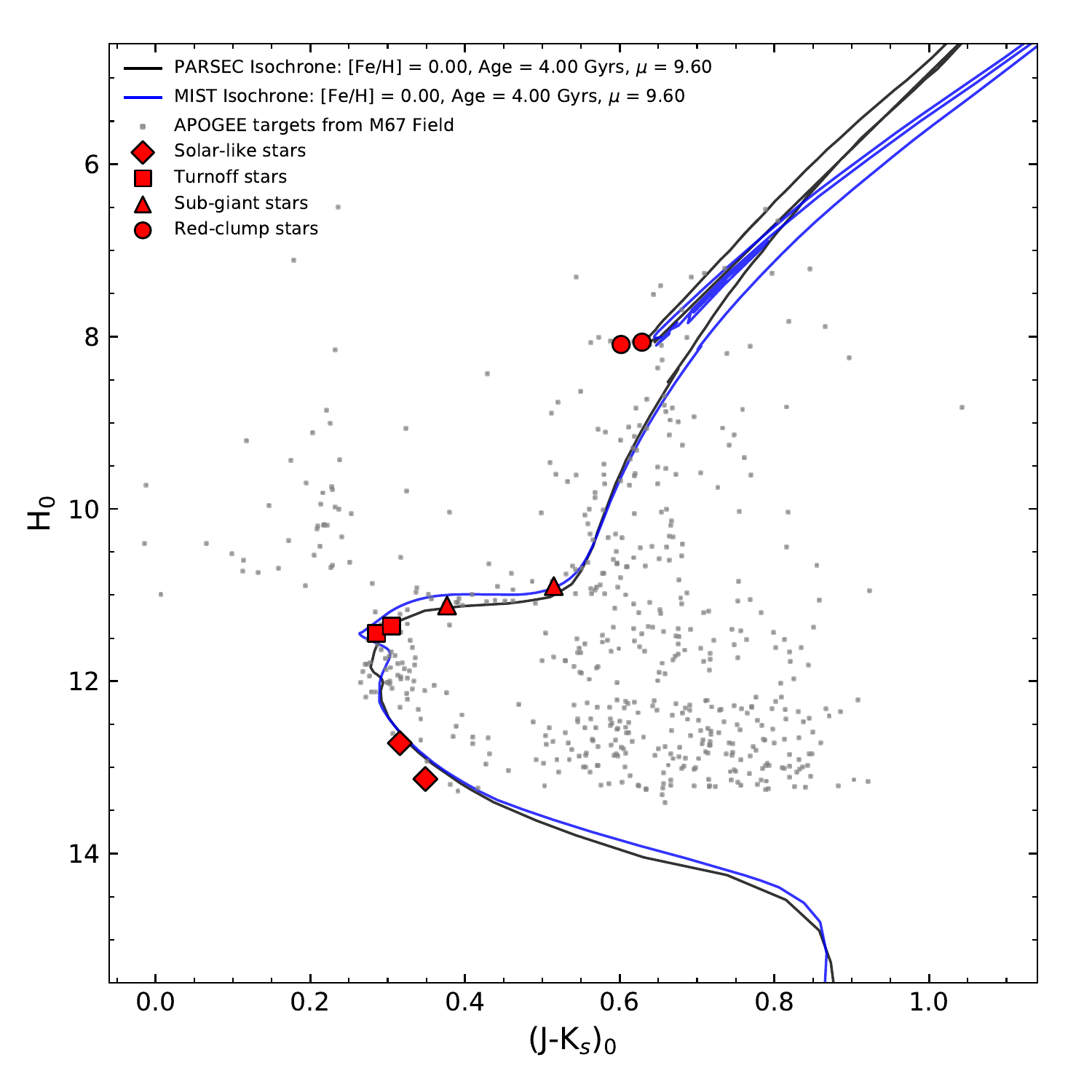}
\caption{2MASS color-magnitude diagram of the APOGEE targets in the M67 field (shown as grey dots). The red symbols correspond to the studied stars: solar-type stars (red diamonds); turnoff stars (red squares); subgiant stars (red triangles); red-clump stars (red circles). Two isochrones for an age of 4 Gyr, (m-M)$_{0}$ = 9.60, and [Fe/H] = 0.00 from PARSEC (black line) and MIST (blue line) are also shown.}
\end{center}
\label{fig1:fig1}
\end{figure*}

\begin{deluxetable}{lcccccccc}
\tabletypesize{\tiny}
\tablecaption{Atmospheric Parameters}
\tablewidth{0pt}
\tablehead{
\colhead{} &
\colhead{2M08510076} &
\colhead{2M08512314} &
\colhead{2M08514122} &
\colhead{2M08505182} &
\colhead{2M08513540} &
\colhead{2M08514474} &
\colhead{2M08521856} &
\colhead{2M08514388} \\  
\colhead{} &
\colhead{+1153115} &
\colhead{+1154049} &
\colhead{+1154290} &
\colhead{+1156559} &
\colhead{+1157564} &
\colhead{+1146460} &
\colhead{+1144263} &
\colhead{+1156425} \\
\colhead{} &
\colhead{G-dwarf} &
\colhead{G-dwarf} &
\colhead{G-turn-off} &
\colhead{G-turn-off} &
\colhead{G-subgiant} &
\colhead{G-subgiant} &
\colhead{K-giant} &
\colhead{K-giant}
}
\startdata
$B$					& 15.498			& 14.804	& 13.338	& 13.271	& 13.500			& 13.848			& 11.427			& 11.579	\\
$V$					& 14.777			& 14.163	& 12.777	& 12.722	& 12.764			& 12.944			& 10.354			& 10.461	\\
$J$					& 13.474			& 13.017	& 11.703	& 11.646	& 11.427			& 11.357			& 8.572				& 8.618		\\
$H$					& 13.157			& 12.741	& 11.466	& 11.382	& 11.143			& 10.918			& 8.087				& 8.114		\\
$Ks$				& 13.105			& 12.681	& 11.397	& 11.321	& 11.030			& 10.822			& 7.923				& 7.996		\\
pm$ra$				& -12.5 $\pm$ 2.5	& -9.5 $\pm$ 1.6	& -8.7 $\pm$ 1.3 & -11.7 $\pm$ 1.3	& -9.7 $\pm$ 1.3	& -10.3 $\pm$ 1.2	& -11.9 $\pm$ 1.5	& -9.7 $\pm$ 2.5\\	
pm$dec$				& -1.9 $\pm$ 2.3	& -1.3 $\pm$ 1.5	& -2.0 $\pm$ 1.3 & -1.8 $\pm$ 1.3	& -2.5 $\pm$ 1.3	& -2.3 $\pm$ 1.2	& -4.9 $\pm$ 1.5	& -1.8 $\pm$ 1.3\\
Probability			& 98				& 99		& 100		& 100		& 91				& 99				& 96				& 97		\\
$RV$				& 33.9 $\pm$ 0.10	& 33.6 $\pm$ 0.10	& 33.6 $\pm$ 0.10	& 30.8 $\pm$ 0.10	& 33.4 $\pm$ 0.10	& 33.1 $\pm$ 0.10	& 33.7 $\pm$ 0.10	& 32.9 $\pm$ 0.1\\
$SNR$				& 120				& 120		& 214		& 210		& 238				& 351				& 504				& 956	\\
$T_{\rm eff}$ (K)	& 5724 $\pm$ 92		& 5958 $\pm$ 33		& 6119 $\pm$ 26		& 6063 $\pm$ 35		& 5596 $\pm$ 38		& 5137 $\pm$ 58		& 4842 $\pm$ 23		& 4819 $\pm$ 82 \\
log $g$				& 4.48 $\pm$ 0.05	& 4.35 $\pm$ 0.05	& 3.91 $\pm$ 0.05	& 3.87 $\pm$ 0.05	& 3.77 $\pm$ 0.05	& 3.64 $\pm$ 0.05	& 2.45 $\pm$ 0.05	& 2.44 $\pm$ 0.05\\
$\xi$ $Km/s$		&	1.00			& 1.00				& 1.15				& 1.20				& 1.25				& 1.20				& 1.75				&1.60		\\
\enddata
\tablewidth{0pt}	
\end{deluxetable}

\section{Chemical Abundance Analysis}

The sample studied here contains a mixture of stellar types.
Both of the K-giants in the sample are found to be red clump (RC) giants by  \cite{Stello2016}, based upon K2 asteroseismology.  
In addition, \cite{Stello2016} analyzed the K2 oscillations from one of the stars, 2M08511474+1146460, and confirmed it to be a subgiant star very close to the base of the RGB.

The spectra of all eight stars were analyzed for chemical abundances in a homogeneous way that is independent from the methodology adopted in the derivation of stellar parameters and chemical abundances for DR14 (14th SDSS Data Release) using the APOGEE automatic abundance pipeline ASPCAP \citep{GarciaPerez2016}.  
The 'boutique', manual abundance analysis described in this section provides independent results for a small sample that can be compared to the automated ASPCAP results for a larger sample of M67 members.

\subsection{Effective Temperatures}

The effective temperatures for the stars in the studied sample were derived using the photometric calibrations of \cite{GonzalezHernandez2009} and five different colors ($B$-$V$, $V$-$J$, $V$-$H$, $V$-$Ks$, and $J$-$Ks$), while adopting a solar metallicity. 
The individual magnitudes $B$ and $V$ were taken from \cite{Yadav2008} and the infrared colors are from the 2MASS catalog \citep{2MASS}. A reddening E(B-V) = 0.041 (\citealt{Taylor2007}; \citealt{Sarajedini2009}) was adopted, with individual dereddened colors obtained using the relations from \cite{Schlegel1998} and \cite{Carpenter2001}. The adopted colors and derived effective temperatures, plus the standard deviations of the mean (typically below $\sim$50 K), are presented in Table 1.
When the internal uncertainties in the \cite{GonzalezHernandez2009} calibration are included, along with the errors in the photometric colors, the total estimated uncertainty expected in $T_{\rm eff}$ is $\sim$100 K, adding all of the errors in quadrature. 

\subsection{Surface gravities}

Stellar surface gravities (log $g$) were derived from the fundamental relation with stellar mass, $T_{\rm eff}$, and absolute bolometric magnitude (Equation 1 below). 
Stellar masses were estimated by using the absolute magnitudes for the $B$, $V$, $H$, $J$, and $Ks$ filters, along with an adopted PARSEC isochrone for an age = 4.0 Gyr, [Fe/H] = 0.0, and (m-M)$_{0}$ = 9.60, with derived masses then being $\sim$ 1.34 $M_{\odot}$ for the red clump, $\sim$ 1.30 $M_{\odot}$ for the subgiants, $\sim$ 1.20 $M_{\odot}$ for the turnoff stars, and $\sim$ 1.00 $M_{\odot}$ for the solar-type stars. 
We note that if the isochrone from MIST (Figure 1) is adopted the obtained stellar masses are not significantly different.
The combination of effective temperature, stellar mass, and bolometric magnitudes (with bolometric corrections from \citealt{Montegriffo1998}) in Equation 1 provides the surface gravity values listed in Table 1.
The solar values used were: log $g_{\odot}$ = 4.438 dex, $T_{\rm eff, \odot}$ = 5772 K and $M_{bol,\odot}$ = 4.75, which follows the IAU prescription in \cite{IAU_solar}.

\begin{equation}
\log{g} = \log{g_{\odot}} + \log\left(\frac{M_{\star}}{M_{\odot}}\right) + 4 \log\left(\frac{T_{\star}}{T_{\odot}}\right) + 0.4(M_{bol,\star} - M_{bol,\odot}),
\end{equation}

Estimates of the uncertainties in the surface gravities were computed from two isochrones, with ages of 3.5 Gyr and 4.5 Gyr, that were used to rederive the stellar masses. 
Included in these estimates were the effective temperature uncertainties, as well as a typical metallicity uncertainty of $\pm$0.05 dex. 
Errors in the photometric magnitudes are small, and of the order of 0.03 mag. Combining all of these values in quadrature, an uncertainty of $\sim$ 0.07 dex is found for the derived values of log $g$. 
Our sample includes three stars with asteroseismic log $g$ from the K2 mission reported in \cite{Stello2016}. Our derived log $g$'s agree quite well with the asteroseismic ones; $\delta$log $g$ (This work - \citealt{Stello2016}) = 0.04 $\pm$ 0.07 dex.

\subsection{Chemical Abundances and Selected Lines}

Chemical abundances were derived from a 1-D LTE analysis and spectral synthesis using the Turbospectrum code (\citealt{Plez2012}, \citealt{AlvarezPLez1998}) in combination with model atmospheres interpolated from the MARCS$\footnote{marcs.astro.uu.se}$ grid \citep{Gustafsson2008} for the atmospheric parameters derived in Sections 3.1 and 3.2 (Table 1). 
The APOGEE line list used in all computations was an updated version of the DR14 line list (line list 20170418). \cite{Shetrone2015} provide details on the construction of the APOGEE line list and updates can be found in Holtzman et al. (\textit{in preparation}).

In a 1-D abundance analysis, the microturbulent velocity ($\xi$) is a necessary parameter to have both weak and strong lines of a given species yield the same abundance. Microturbulent velocities were determined using the same procedure as in \cite{Souto2016} (or \citealt{Smith2013}).
The determination of $\xi$ relies on Fe I lines that span a range of line strengths (or equivalent widths), with the stronger lines displaying a much larger sensitivity of the derived abundance with the microturbulent velocity.
The best value of $\xi$ yields the closest agreement in the abundances of the strong and weak lines. In practice, adopted values of $\xi$ were varied from 0.5 to 3.0 km s$^{-1}$. 
The inferred values of $\xi$ are included in Table 1.

Chemical abundances of individual elements were derived using a line-by-line manual analysis and obtaining best fits of synthetic spectra to observed line profiles. Local continuum levels in the observed spectra were set manually, and the particular elemental abundance varied, until differences between observed and sythetic spectra were minimized.
The instrumental resolution of the APOGEE spectrograph ($R$ $\sim$ 22,500) produces an instrumental profile with a $\sim$13.7 km s$^{-1}$ full-width-half-maximum (FWHM $\sim$ 0.71 \AA{}). Small variations in the broadening result in small adjustments across the spectra about $\sim$$\pm$1.5 km-s$^{-1}$.  
During the line-profile fitting, searches were made for extra-broadening effects related to v sin($i$) and/or macroturbulence; however, no extra line-broadening was needed, beyond the instrumental APOGEE profile to obtain good fits to the observed line profiles.

Chemical abundances of the elements C, N, O, Na, Mg, Al, Si, K, Ca, Ti, V, Cr, Mn, and Fe, were derived for the two red clump (RC) giants in our sample (Table 1). 
The selected transitions were the same as in our previous study of red giants in the open cluster NGC 2420 using APOGEE spectra (\citealt{Souto2016}). \cite{Souto2016} analyzed a sample of red giants with similar values of $T_{\rm eff}$ and log $g$ (although slightly more metal-poor, with [Fe/H]=-0.20) to the M67 RC giants studied here. 

The previous work on NGC 2420 by \cite{Souto2016} did not analyze solar-type dwarfs, turnoff stars, or subgiants. In this study, we have made a careful search for usable spectral lines or features in the APOGEE spectra of dwarfs and subgiants, with the goal being to maximize the number of lines available for each chemical species.  
Initial identification of promising spectral lines in solar-type dwarfs were done using an APOGEE spectrum of the asteroid Vesta (as a solar proxy), which were observed using the APOGEE spectrograph fiber-linked to the APO 1-m telescope. 

The atomic and molecular lines used in the ``boutique'' abundance analysis of the stars (and their associated abundances) are listed in Table 2. 
A total of 135 spectral lines or features were selected as abundance 
indicators: 77 Fe I, 4 CO, 3 C I, 10 CN, 4 OH, 2 Na I, 6 Mg I, 3 Al I, 9 Si I, 2 K I, 4 Ca I, 6 Ti I, 1 V I, 1 Cr I, and 3 Mn I. 
As the studied sample covers an extended range in $T_{\rm eff}$-log $g$ parameter space, it is not possible to measure the exact same transitions for all stars, given that the strengths of the various spectral features change as a function of $T_{\rm eff}$ and log $g$. 

The APOGEE spectra of cool red giants are typically dominated by molecular features (mostly CO, CN, and OH), but also show atomic lines from many elements. For G-dwarfs and subgiants, with higher effective temperatures, molecular absorption becomes less important and neutral atomic lines become dominant. 
In the case of G-dwarfs and turnoff stars, it is not possible to derive oxygen and nitrogen abundances as the molecular OH and CN lines become too weak.
Vanadium abundances are also not measurable.
Most chemical abundances from APOGEE are based upon transitions of neutral atomic lines and include Fe, Na, Al, Mg, Si, K, Ca, Ti, V, Cr, and Mn. These lines can be measured, for the most part, in all four stellar classes, although for some lines the degree of blending and/or the line strengths change significantly, depending on the stellar class.
Concerning iron, often taken as a metallicity indicator, a relatively small number of clean Fe I lines can be analyzed in red giants, while more than 70 Fe I lines can be used in the warmer subgiants, turnoff stars, and G-dwarfs.
Although the CO lines become quite weak for solar-type stars, three C I lines ($\lambda$15784.7\AA{}, $\lambda$16005.0\AA{}, $\lambda$16021.7\AA{}) become stronger and measurable in these stars. 

Figure 2 illustrates the observed and best-fit synthetic spectra for one star in each class: one red clump, one subgiant, one turnoff star, and one solar-type G-dwarf (from top to bottom).
These spectra highlight the region between $\lambda$16140\AA{} and $\lambda$16270\AA{}, covering only a small portion of the APOGEE wavelength range. 
Many of the spectral lines in Figure 2 become noticeably broader in the G-dwarf spectra due, primarily, to increasing surface gravity.  
The nature of the absorption lines change noticeably when going from the K-giant (top panel) down to the G-dwarf (bottom panel), with the weakening of the molecular lines. 

\begin{figure*}
\figurenum{2}
\begin{center}
\includegraphics[width=1\linewidth]{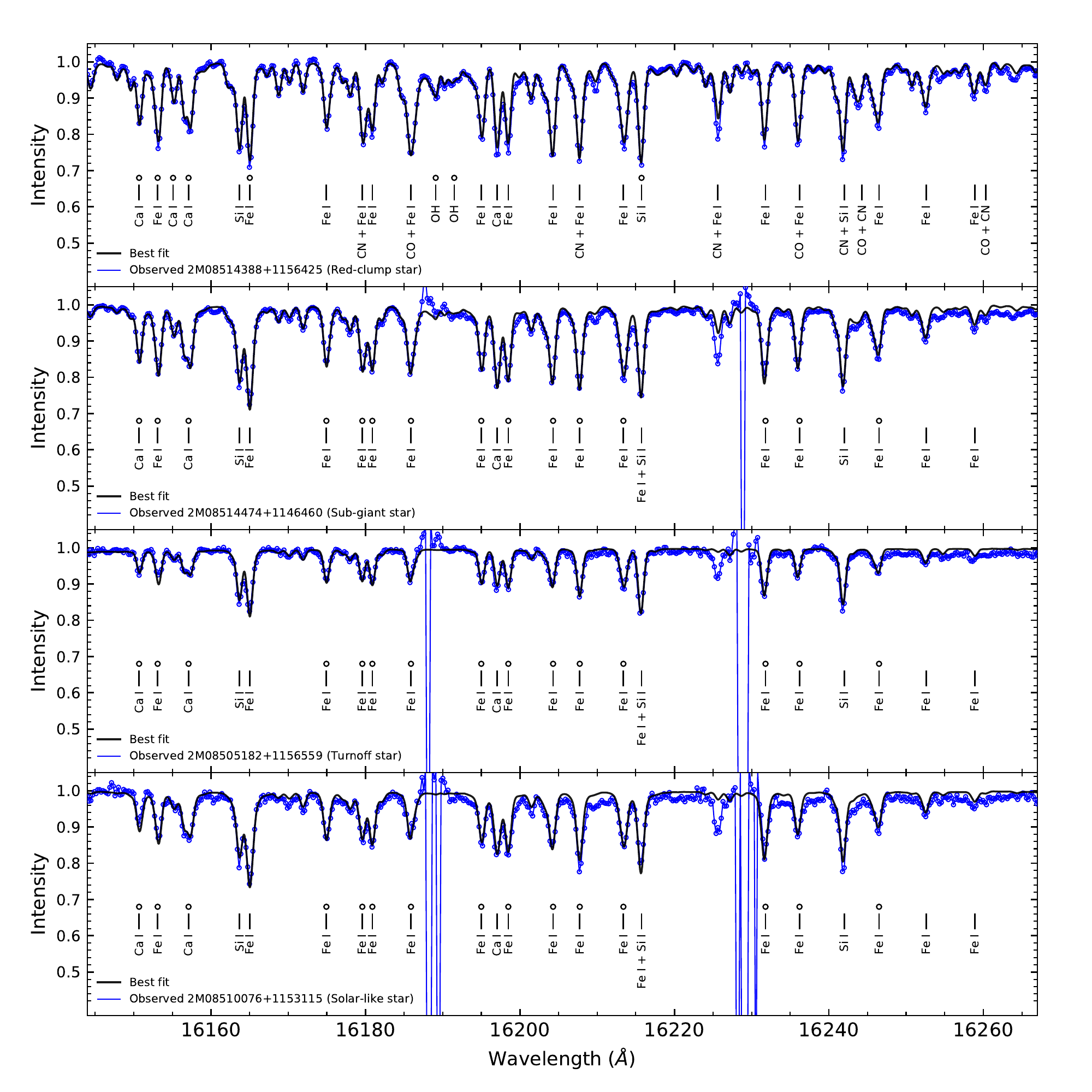}
\caption{Best fitted synthetic spectra (in black) overploted with the observed spectra (in blue) for four target stars: one red clump, one subgiant, one turnoff, and one solar-type star, respectively, from top panel to bottom panel. The spectral lines used to derive abundances are marked with black circles.}
\end{center}
\label{fig2:fig2}
\end{figure*}

Abundance results from line-by-line measurements are shown in Table 2, while mean elemental abundances obtained for each star are presented in Table 3.

\subsection{Abundance Uncertainties}

To estimate the uncertainties in the derived abundances due to uncertainties in the adopted stellar parameters, new abundances were computed for perturbed values of the microturbulent velocity, as well as using model atmospheres with perturbed values of effective temperature, surface gravity, and metallicity.

For all stellar classes, the baseline models corresponded to the stars: 2M08510076+1153115 (solar-type), 2M08505182+1156559 (turnoff), 2M08513540+1157564 (subgiant), and 2M08521856+1144263 (red clump). In each case, $T_{\rm eff}$ was changed by +100 K, log $g$ by +0.20 dex, [Fe/H] by +0.20 dex, and $\xi$ by +0.20 km s$^{-1}$. Table 4 presents the final estimated abundances uncertainties, $\sigma$, representative of each stellar class. The final uncertainties were computed from the sum in quadrature of all the estimated uncertainties (same procedure as \citealt{Souto2016,Souto2017}).

The final abundance uncertainties for all stars are overall similar. The abundances of Mg, Al, and Si are more sensitive to changes in both $T_{\rm eff}$ and log $g$; their uncertainties are $\sim$ 0.10 dex. 
In solar-type stars, the most sensitive elements to change in atmospheric parameters are Mg and Al, while for subgiants the abundances of Mg and Si exhibit higher sensitivity to changes in the atmospheric parameters. 
The abundances of red clump indicate a larger dependence to changes in $\xi$ for Fe I and OH lines. Titanium is found to have highest sensitivity to $T_{\rm eff}$, while the oxygen abundance from OH lines is found to be more sensitive to changes in [Fe/H]. Variations of log $g$ by +0.20 dex in the model do not change significantly the abundances in red giants.

\subsection{Comparisons with Optical Studies from the Literature}

Figures 3 and 4 show comparisons of the abundances obtained for all elements studied here with results from M67 high-resolution optical studies in the literature by \citeauthor{Liu2016} (2016; solar twins), \citeauthor{Onehag2014} (2014; solar-type, turnoff and subgiant stars), and the studies of red giants by \cite{Tautvaisiene2000}, \cite{Yong2005}, \cite{Friel2010}, and \cite{Pancino2010}. All abundances are plotted as a function of log $g$ (Figure 3) and $T_{\rm eff}$ (Figure 4), with the [X/H] values from this study using our solar results as a reference, while  [X/H] values from the literature were taken directly from the other studies using their solar abundances as references. We note that these literature studies have analyzed the solar spectrum themselves (except for \citealt{Onehag2014}) and they use their derived solar abundances to measure differential abundances relative to the Sun. Our solar abundance results are also presented in Table 2 and 3. 

\begin{figure*}
\figurenum{3}
\begin{center}
\includegraphics[width=1\linewidth]{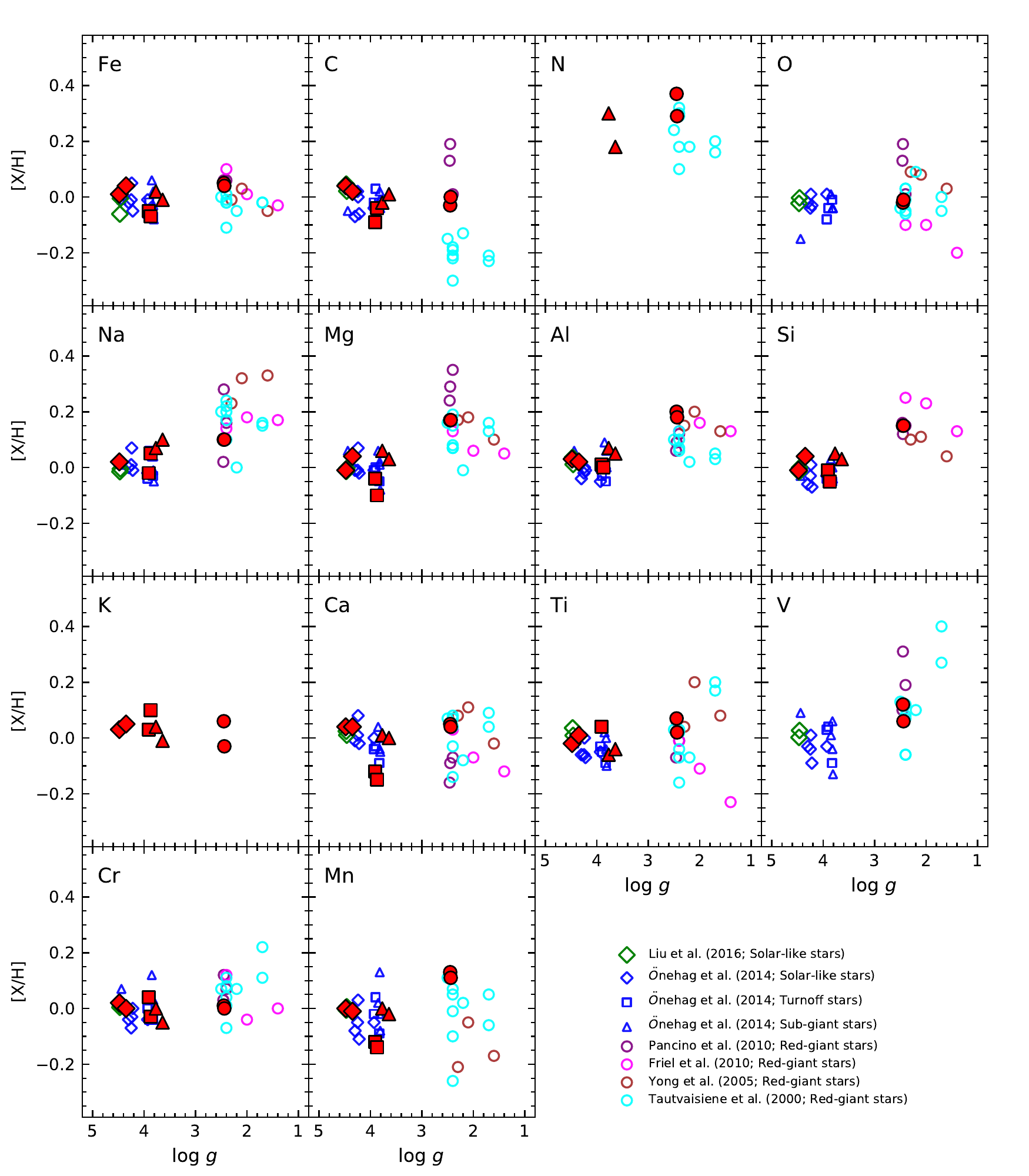}
\caption{The chemical abundances for the studied stars are shown as a function of log $g$. Results from the literature for other stars in M67 from \cite{Liu2016}, \cite{Onehag2014}, \cite{Pancino2010}, \cite{Friel2010}, \cite{Yong2005}, and \cite{Tautvaisiene2000} are also show for comparison. The results from this study are shown as filled red symbols as in Figure 1.}
\end{center}
\label{fig3:fig3}
\end{figure*}

\begin{figure*}
\figurenum{4}
\begin{center}
\includegraphics[width=1\linewidth]{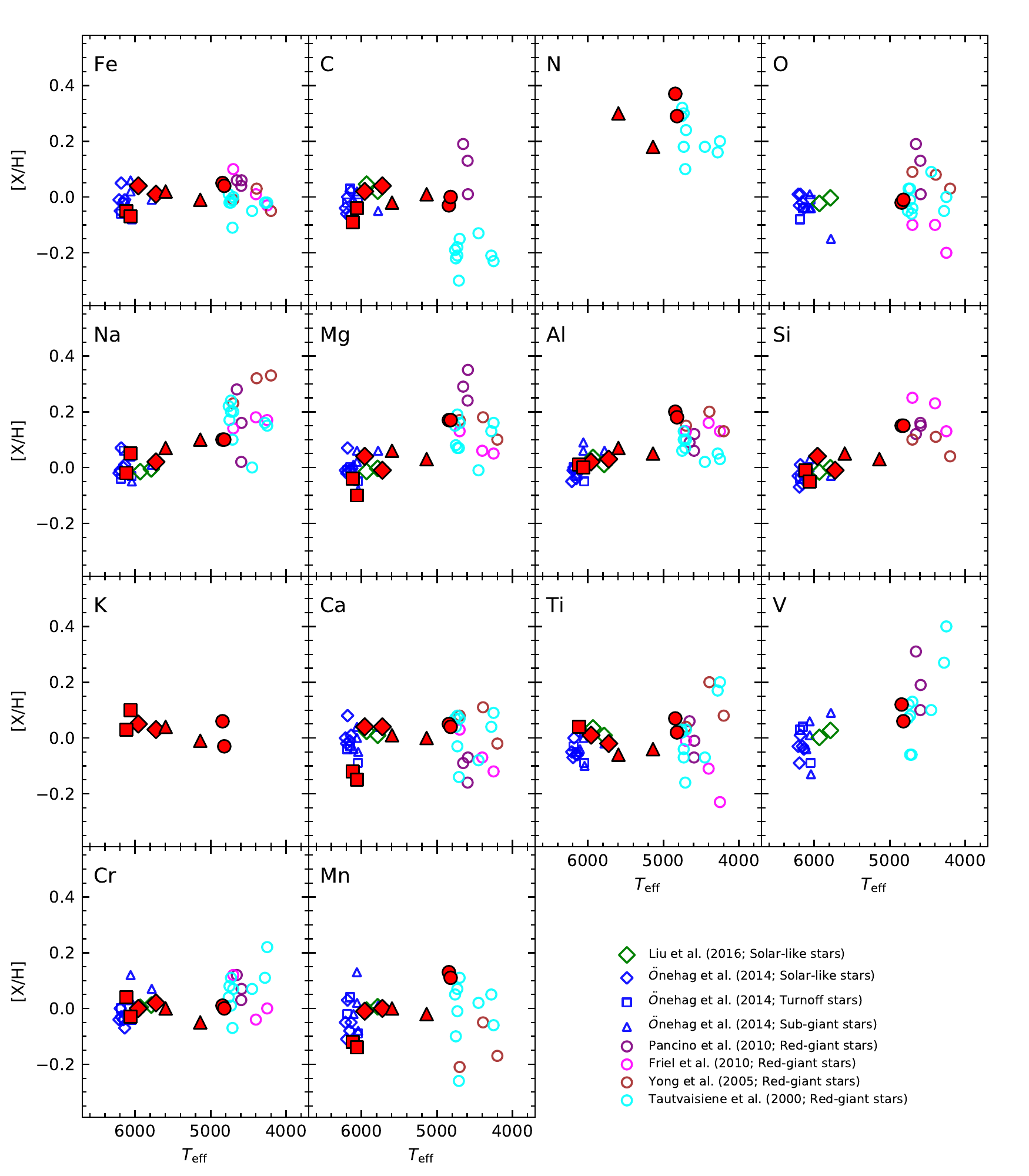}
\caption{Same as Figure 3, except shown as function of $T_{\rm eff}$.}
\end{center}
\label{fig4:fig4}
\end{figure*}

Inspection of Figures 3 and 4 show that our abundances for solar-type stars in M67 are in overall good agreement with those from both \cite{Liu2016} and \cite{Onehag2014}.
For the turnoff stars we obtain results that are also similar to the ones from \cite{Onehag2014}; our abundances of Mg, Ca, and Mn are systematically lower by roughly 0.05 -- 0.10 dex. 
For the subgiant stars our results agree in general with \cite{Onehag2014}; the largest offset is found for Na, where we obtain an average Na abundance 0.09 dex higher than in \cite{Onehag2014}.

Several studies have determined chemical abundances of red giants in M67 via high-resolution spectroscopy and a few of these works are compared in Figures 3 and 4. 
Overall our red clump abundances for all elements are similar to the literature values, falling close to the middle of the distribution of the abundances for the red clump in Figures 3 and 4, except for the elements N, Al, Ca, and Mn, (and to a certain degree Ti), for which our results fall in the upper envelope of the distribution by $\sim$ 0.05 -- 0.20 dex. We also note significant abundance scatter in the optical literature results for O, Na, Ca, Ti, V, and Mn.

It is also noted that the abundances of Mg, Al, and Si in red giants are found to be, on average, enhanced relative to solar.
Concerning metallicities, the mean iron abundances obtained for the two red clump stars studied here is $\langle$[Fe/H]$\rangle$ = +0.05 and these results agree with \cite{Pancino2010}, but are slightly higher than the average metallicity from \citeauthor{Tautvaisiene2000} (2000;  $\langle$[Fe/H]$\rangle$ = -0.03).
Overall, the abundance patterns derived from the literature studies are similar to those obtained from the APOGEE spectra (with some significant offsets or scatter in certain elements). It is worth noting here again that all studies, this work included, are based upon 1-D LTE analyses.

\section{Abundances Trends}

Chemical abundance trends may result from a combination of simplification in the analysis or possible physical effects, such as non-LTE, 3-D, and/or diffusion, as well as systematic errors in the abundance analysis.
This subsection will discuss observed abundance trends across stars in different evolutionary phases for the abundance results presented in this study.  
This discussion will not include C and N, as abundance changes due to the first dredge-up dominate any other physical processes, such as diffusion (Section 5.1).
In addition, O and V are not included, as abundances from these elements are only measurable in the APOGEE spectra for red giants. 
This leaves 10 elements to investigate as a function of evolutionary state; simply put, stellar evolution proceeds monotonically in log $g$ from high log $g$ to decreasing log $g$ as the star evolves to the RGB phase. 

Focusing on the abundance results obtained here as a function of log $g$ (red filled symbols in Figure 3), two main types of behavior can be noted: in one case there is almost no change in the abundances as a function of surface gravity, which occurs for K, Ti, and Cr.
In the other case, there are general increases in the abundances when comparing the main sequence, turnoff, and subgiant stars with the red clump stars (decreasing log $g$). Among these 7 elements, 5 have larger trends than the other 2 when comparing the less evolved stars and the red clump stars. 
These larger trends with surface gravity ($\sim$0.20 dex) are found for the abundances of the elements Mg, Al, Si, Ca, and Mn, with the mean abundances of the turnoff and red clump stars being:
$\langle$A(Mg)$\rangle$ = 7.37 and $\langle$A(Mg)$\rangle$ = 7.61;
$\langle$A(Al)$\rangle$ = 6.38 and $\langle$A(Al)$\rangle$ = 6.56;
$\langle$A(Si)$\rangle$ = 7.44 and $\langle$A(Si)$\rangle$ = 7.62;
$\langle$A(Ca)$\rangle$ = 6.20 and $\langle$A(Ca)$\rangle$ = 6.38;
$\langle$A(Mn)$\rangle$ = 5.27 and $\langle$A(Mn)$\rangle$ = 5.52 for turnoff and red-clump stars, respectively. 
The abundances of Na and Fe also show a trend with log $g$, but with a smaller change  ($\delta$$\sim$0.1 dex). 
For sodium and iron, we obtain: 
$\langle$A(Na)$\rangle$ = 6.33 and $\langle$A(Na)$\rangle$ = 6.41;
$\langle$A(Fe)$\rangle$ = 7.39 and $\langle$A(Fe)$\rangle$ = 7.50, for the turnoff and red clump, respectively.
Three of these elements, Na, Al, and Si, display the smallest differences between the abundances of the turnoff and solar-type stars and show a more monotonic increase with decreasing log $g$.

Such trends in the abundances as a function of log $g$ are also found when examining the literature
optical results (shown in Figure 3), indicating behavior reminiscent of that found here for the APOGEE results. For example, \cite{Liu2016} studied solar-twins in M67 and derived [Na/H] = -0.05 $\pm$ 0.02, while the study of red giants by \cite{Friel2010} obtained [Na/H] = +0.11 $\pm$ 0.10.  
The Al abundance for the solar twins from \cite{Liu2016} is [Al/H] = +0.01 $\pm$ 0.03, while for the red giants \cite{Friel2010} obtain [Al/H] = +0.08 $\pm$ 0.07. 
Taking the Mg abundances as a second example, and the literature results from \cite{Onehag2014} for the turnoff stars and \cite{Pancino2010} for the red giants, we see a change in the abundances of 0.31 dex in the mean. 

Overall, the abundance results for most of the elements from all studies, taken at face value, would suggest that the abundances may increase as the star evolves (not expected), or alternatively, that other effects become significant in the red giant regime. However, it must also be kept in mind that comparisons of results from different studies using different techniques and analysis methods may result in systematic differences in the abundances. 

The distribution of the elemental abundances as a function of the effective temperature (Figure 4) can also be divided in elements that show small to no trends with $T_{\rm eff}$ (such as, K, Ti, and Cr) and those that show a trend of increasing abundances with $T_{\rm eff}$ between the solar-like and the K-type red giants (which may be related to surface gravity): Mg, Al, Si, and Mn with the
largest trends, while Na and Fe display a smaller effect. Calcium, again, shows a change in abundance which is driven exclusively by the low abundances in the turnoff stars.
The abundances of Mg, Ca, Mn, Fe, and to a lesser degree Si, show a pronounced decrease in the abundances of the turnoff stars (higher $T_{\rm eff}$) when compared to the abundances of the solar type and subgiant stars.  

\subsection{Departures from LTE}

The trends of elemental abundances with surface gravity and effective temperature seen for some elements in Figures 3 and 4, respectively, can be due to a combination of effects that include, but are not limited to, abundance offsets due to departures from LTE.
This is a possible effect for the abundances in this study, as the targets cover a large range in $T_{\rm eff}$-log $g$ parameter space, although M67 has solar metallicity and departures from LTE are expected to be more significant for red giants in the metal-poor regime (e.g., \citealt{Asplund2005nonLTE,Asplund2009}). 
Several studies in the literature have investigated non-LTE effects for lines in the optical (e.g., \citealt{Andrievsky2008}, \citealt{Korn2007}, \citealt{Lind2011}, \citealt{Bergemann2012}, \citealt{Smiljanic2016}, \citealt{Osorio2016}); however, to date, few non-LTE studies have investigated the behavior of transitions in the $H$-band and, in particular, in the APOGEE region. 

\cite{Cunha2015} presented non-LTE abundance corrections for Na I lines in the APOGEE region for stars in the red clump and on the red giant branch in the very metal-rich ([Fe/H]=+0.35) open cluster NGC 6791; the departures from LTE were found to be minimal. In this paper, the differences in A(Na) between the red giants (RG) and the solar-like (SL) are $\Delta$(RG - SL)= +0.08 dex. Non-LTE effects are expected to be smaller than these differences and of the order of $\sim$0.02 dex (\citealt{Lind2011}; private communication). 

Two recent studies investigated the formation of Mg I lines \citep{Zhang2017} and Si I lines \citep{Zhang2016} in the APOGEE region. \cite{Zhang2017} found that the Mg I lines $\lambda$ 15740\AA{}, $\lambda$ 15748\AA{}, $\lambda$ 15765\AA{} are well-modeled in LTE (showing only small non-LTE departures). We find that the Mg abundances in the red clump stars are 0.24 dex larger than in the turnoff stars. Such offsets between the abundances are unlikely to be explained as due to departures from LTE. 

The results in \cite{Zhang2016} indicated that the Si I lines analyzed here at $\lambda$15888\AA{}, $\lambda$16380\AA{}, $\lambda$16680\AA{}, and $\lambda$16828\AA{} show departures from LTE: $\delta$A(Si)(non-LTE - LTE) = -0.06 for K-type red-giant stars, $\delta$A(Si)(non-LTE - LTE) = -0.05 and -0.03 for G-type subgiants and solar type-stars, respectively. These non-LTE corrections are all in the same sense (all negative) and the offsets are roughly the same for red giants and subgiants, and slightly smaller (by 0.03 dex) for solar-type stars. These non-LTE corrections of $\sim$0.01--0.03 dex, although reducing the discrepancy, would not seem to be a plausible explanation for the 0.18 dex differences found here between the Si abundances of the red giants and turnoff stars. 

\section{Discussion}

\subsection{Carbon and Nitrogen Abundances -- FDU Signature}

The carbon and nitrogen abundances (Table 2) reveal signs of the first dredge-up (FDU) when comparing the results obtained for subgiants and red clump stars. As discussed previously, the determination of the nitrogen abundances are only possible for the subgiants and giants. 
The signature of the first dredge-up is most apparent in the comparison of the ratio $^{12}$C/$^{14}$N. 
We find: $^{12}$C/$^{14}$N = 2.34 for the subgiants, and $^{12}$C/$^{14}$N = 1.73 for the red clumps. The decrease of this ratio in the red clump stars is indicative of the dredge-up of $^{14}$N that is driven by H-burning during the CN-cycle. 
Concerning oxygen, the OH lines become too weak to be useful to measure the oxygen abundances in stars with $T_{\rm eff}$ $>$ 5000 K. In the case of red clumps, however, an oxygen abundance is found with  $\langle$A(O)$\rangle$ = 8.65 $\pm$ 0.01, consistent with the adopted solar oxygen abundance.

\subsection{Abundance Variations in M67}

The chemical abundances obtained for the eight M67 members studied here, taken at face value, would indicate a measurable abundance spread within M67 for some of the studied elements. However, in all cases the abundances in the two targets of the same stellar class are found to be quite homogeneous, hinting that when comparing stars across significantly different $T_{\rm eff}$--log regimes, the analyses may be detecting effects other than simple primordial abundance dispersions. 

\subsubsection{Solar-type Stars}

The two G-dwarfs studied here have atmospheric parameters similar to that of the Sun ($T_{\rm eff}$ = 5724 K, log $g$ = 4.48 for 2M08510076+1153115; $T_{\rm eff}$ = 5958 K, log $g$ = 4.35 for 2M08512314+1154049). 
The chemical similarities between solar-type stars in M67 and the Sun have been discussed previously  \citep{Onehag2014}.

The elemental abundances of the two G-dwarfs are found to be very consistent with each other, with a mean difference of +0.01 $\pm$ 0.03 dex (in the sense of hotter minus cooler star). 
Figure 5 illustrates the close match to a solar abundance pattern (within less than 0.05 dex) of both solar-like stars in M67. One of the solar-like stars (2M08510076+1153115 or YBP 1514) has an exoplanet detected by \cite{Brucalassi2014,Brucalassi2016} with a minimum M$_{\rm Jup} = $0.40.  
The other solar-like star (2M08512314+1154049 or YBP 1587) has been reported in \cite{Pasquini2012} to be an exoplanet host candidate.
Our derived stellar parameters and metallicity for 2M08510076+113115 suggest that this star is a solar-twin, exhibiting abundance differences relative to the Sun of $\leq$ 0.04 dex for all elements.

\begin{figure*}
\figurenum{5}
\begin{center}
\includegraphics[width=1\linewidth]{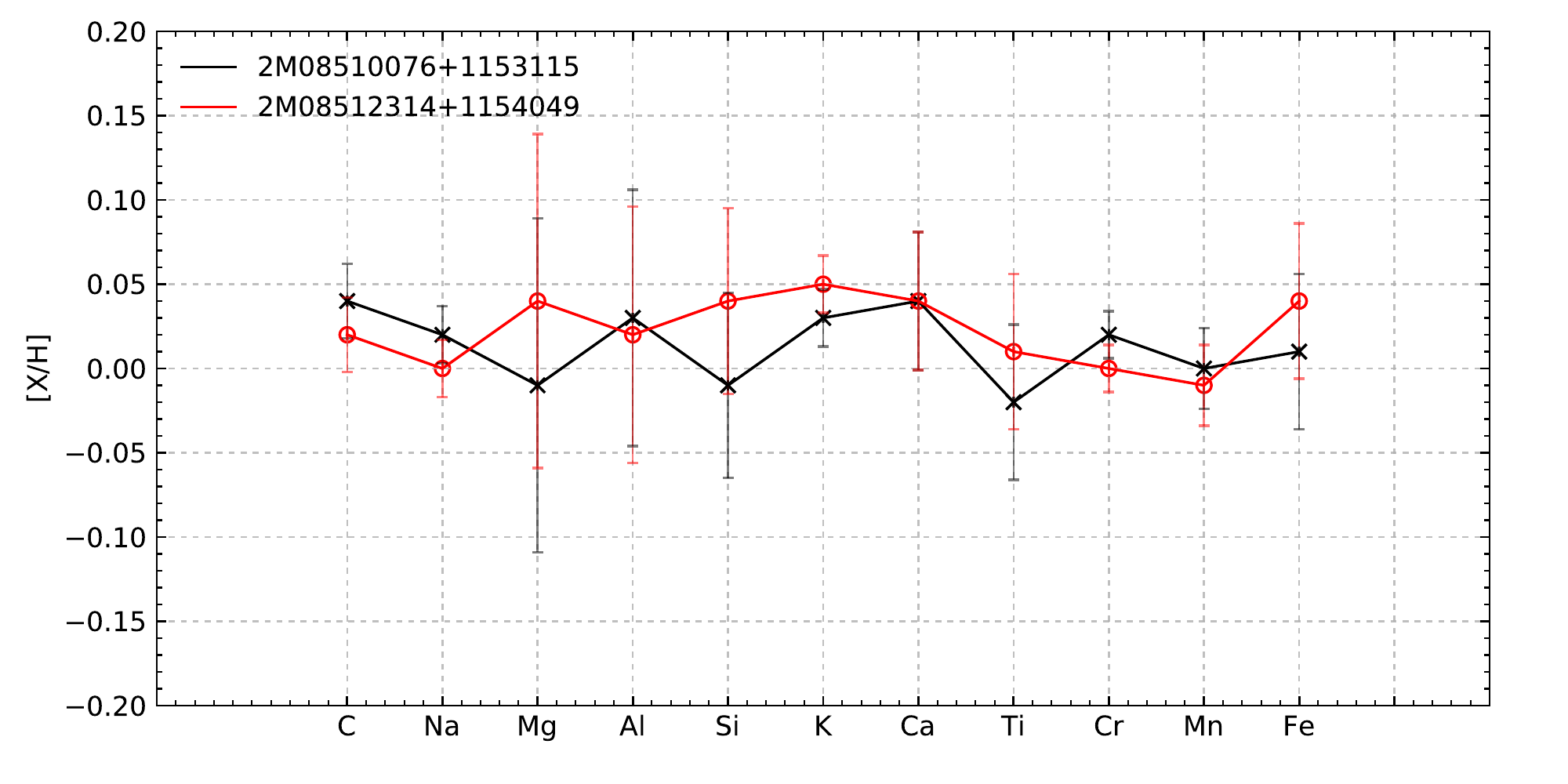}
\caption{The chemical abundances for the two G-type dwarfs relative to Sun ([X/H], with solar results from this study). The solar abundances ([X/H]=0) $\pm$ 0.05 dex are indicated as dashed lines. This illustrates the close match to a solar abundance pattern (within less than 0.05 dex) of both G-dwarfs studied in M67.}
\end{center}
\label{fig5:fig5}
\end{figure*}

Certain stellar abundance ratios are important for studies
of exoplanet properties, such as the C/O ratio, which is one of the factors that affects the ice chemistry in protoplanetary disks (\citealt{Bond2010}; \citealt{Teske2014}). Unfortunately, the APOGEE spectra cannot be used to determine precise oxygen abundances in solar-type stars; however, another interesting abundance ratio that affects the 
interior structure of rocky exoplanets is Mg/Si (\citealt{DelgadoMena2010}; \citealt{BrewerFischer2016}; and \citealt{UnterbornPanero2017}). We derive Mg/Si = 0.93 for both M67 G-dwarf stars$\footnote{Mg/Si = N(Mg)/N(Si) = 10$^{\rm log N(Mg)}$/10$^{\rm log N(Si)}$}$, compared to our same value of 0.93 obtained for the Sun. 

\subsubsection{Turnoff Stars}

We analyzed two stars from the turnoff point of the HR diagram (2M08514122+1154290; $T_{\rm eff}$ = 6119 K, log $g$ = 3.91, and 2M08505182+1156559; $T_{\rm eff}$ = 6063 K, log $g$ = 3.87); the abundances of these two stars have a difference $\delta$ = 0.07 dex for Mg, K, and Cr, but these are still within the uncertainties in the abundance determinations, see Table 4.

\subsubsection{Subgiant Stars}

The subgiants analyzed in this work are 2M08513540+1157564 ($T_{\rm eff}$ = 5596 K, log $g$ = 3.77)  and 2M08514474+1146460 ($T_{\rm eff}$ = 5137 K, log $g$ = 3.64); 2M08513540+1157564 is located between the turn-off of the main sequence and the base of the red-giant branch (RGB), while 2M08514474+1146460 is near the base of the red-giant branch. 
Their carbon and nitrogen abundances are found to be slightly different in the two stars, and such differences are expected based on stellar evolution models (e.g., \citealt{Lagarde2012}): the mean difference ($\pm$ rms) between both subgiant stars are 0.02 $\pm$ 0.04 dex.

\subsubsection{Red-clump Stars}

The red clump stars analyzed here have similar atmospheric parameters: 2M08521856+1144263 with $T_{\rm eff}$ = 4842 K, log $g$ = 2.45 and 2M08514388+1156425 with $T_{\rm eff}$ = 4819 K, log $g$ = 2.45. Both stars are members of the red clump (RC) of M67.
As with the other pairs of stellar types isolated here, these M67 red clumps share a nearly identical chemistry, with the mean difference in elemental abundances being
+0.02 $\pm$ 0.03 dex (where the difference is hotter giant - cooler giant).  
The largest difference is 0.08 dex for $^{14}$N, which could be the result of slightly different mixing and mass loss histories through the first dredge-up and He core flash.

\subsection{Signatures of Diffusion in M67}

Within the pairs of stars of the same stellar classes analyzed here, their chemical compositions are quite homogeneous, while a comparison across the stellar classes may be used to probe the existence and extent of diffusion.
Convective mixing predicts (e.g., \citealt{Lagarde2012}) that red-giant photospheres become richer in nitrogen, due to internal stellar nucleosynthesis and deep mixing; however, an increase in the abundances of elements such as Na, Mg, Al, Si, Ca, Mn, and Fe is not expected in low-mass giants, such as those found in M67. 
Small increases in red giant abundances, relative to main-sequence and perhaps subgiant stars might be associated with stellar diffusion operating in the hotter main-sequence stars, while the convective envelopes developing in the atmospheres of evolved stars would tend to erase the diffusion signature. As pointed out in \cite{Dotter2017}, atomic diffusion mechanism operates most effectively in the stars' radiative regions.

\cite{Onehag2014} investigated possible diffusion signatures in M67 by studying a sample of hot, main sequence stars just below the turnoff ($T_{\rm eff}$$\sim$6130K - 6200K), turnoff stars ($T_{\rm eff}$$\sim$6150K - 6215K), and early subgiant stars ($T_{\rm eff}$$\sim$6040K - 6110K).  
Comparing the abundances in the subgiant and turnoff stars, \cite{Onehag2014} found differences in elemental abundances of
$\Delta$(SG - TO)$\sim$+0.02 -- +0.06 dex; this is in the correct sense predicted by models of diffusion, as the heavy elements that have sunk below the small convection zones in the hotter turnoff stars are mixed back to the observable surface by the deepening convective envelopes in the subgiants.  
The specific elements studied by \cite{Onehag2014} were C, O, Na, Al, Si, Mg, S, Ca, Ti, Cr, Mn, Fe, and Ni, and our study includes 10 of these elements in the G-dwarfs and subgiants.

Results derived here for the turnoff, subgiant and red clump stars, compared to the solar twin (2M08510076+1153115), are examined to search for the effects of diffusion in M67.  
Figure 6 plots differences in the abundances between the turnoff, subgiants, and red clumps (their mean abundances for each class) minus the solar twin.  
The red clump stars display larger abundances in most elements relative to the solar twin which may reflect the convective erasure of a diffusion signature in the solar twin and even more to the turnoff stars, or the pattern may point to systematic effects in the analysis, as the two types of stars have quite different values of effective temperature and surface gravity (see Section 3).

The abundance differences in Figure 6 between the subgiants and the solar twin are much smaller in comparison with the red clumps, with differences typically $<$0.05 dex.  
The stellar parameters between these stars are much more similar than for the red clump stars.
The pattern in the differences is interesting and warrants closer examination.
The magnitude of diffusion varies from element-to-element and \cite{Onehag2014} also indicate in their Figure 8 that the expected magnitude of the effect of diffusion for each element in M67 stars on the warm main sequence and turnoff.
From \cite{Onehag2014}, the order of the magnitude of diffusion differences would be expected to be the largest in Na, Mg, Al, Fe, and C, while differences in Mn, Cr, and Si would be smaller, and for Ca and Ti almost non-existent.  
Examining the subgiant - solar twin pattern, we note that Na, Mg, and Al all exhibit relatively strong positive differences, as does Si, Ca, Mn, and Fe.  
Interestingly, Ca, and Cr show small negative differences.  
The derived abundances for the turnoff stars set the lower limits for elemental abundances in M67. We obtain for almost all species abundances differences $<$ 0.10 dex compared to the solar twin as well as for the other classes. The exceptions were abundances from K, Ti, and Cr, the elements that showed less change in their abundances as a function of log $g$.
Given that there may be small, systematic effects in the absolute abundance scale as derived for the subgiants, turnoff, and the solar twin (a few hundredths of a dex), the overall pattern in Figure 6 may indicate that the signature of diffusion in M67 is stronger in turnoff stars, followed by solar like stars.

\begin{figure*}
\figurenum{6}
\begin{center}
\includegraphics[width=1\linewidth]{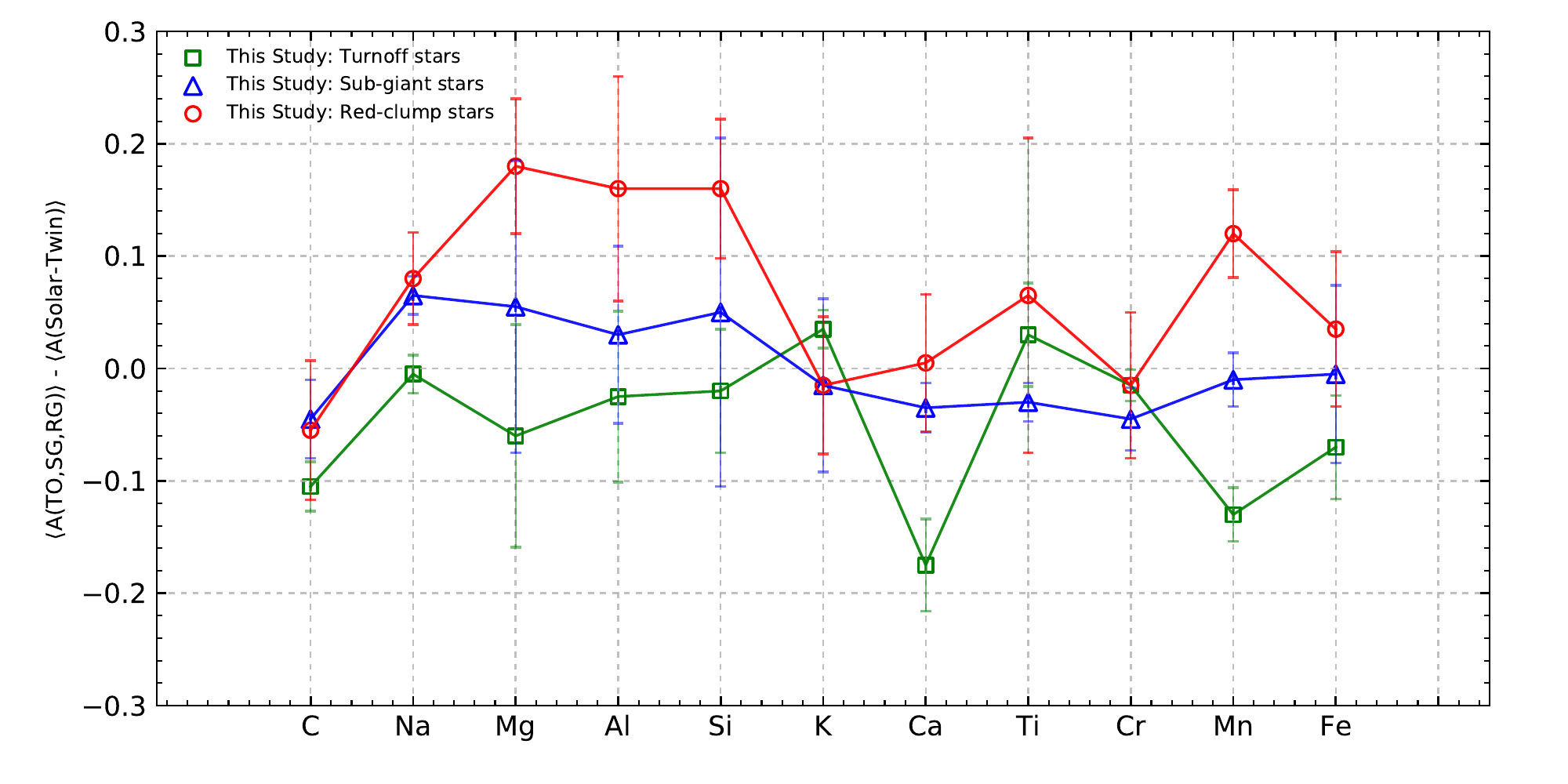}
\caption{Mean abundances for red clump (red curve), subgiants (blue curve), and turnoff (green curve) minus the abundances for the solar twin 2M08510076+1153115. The red clump and turnoff stars show systematically higher and lower elemental abundances, respectively, compared to the solar twin.}
\end{center}
\label{fig6:fig6}
\end{figure*}
Figure 6 also indicates that we have an increasing overall metallicity from the turnoff stars to the red-clump, with the subgiants and the solar-like stars marginally showing similar levels of chemical compositions. 
When computing overall metallicities for each class (using the [M/H] as the sum of all elemental abundances) we obtain: for the solar-like stars $\langle$[M/H]$\rangle$ = 0.03 $\pm$ 0.05, turnoff stars $\langle$[M/H]$\rangle$ = -0.02 $\pm$ 0.00, subgiants stars $\langle$[M/H]$\rangle$ = 0.02 $\pm$ 0.07, and for the red-clump $\langle$[M/H]$\rangle$ = 0.09 $\pm$ 0.07, where the uncertainty here represents the standard deviation$\footnote{The uncertainty in the mean is a factor 1/sqrt(2) smaller.}$ for all elements shown in Figure 6.

Recently, \cite{Dotter2017} published predicted surface abundance changes in stellar models that were computed with the MESA code and included atomic diffusion (\citealt{Paxton2011,Paxton2013,Paxton2015}, see also \citealt{Dotter2016} and \citealt{Choi2016}).
Surface abundances of a number of elements were presented for stars in different evolutionary stages (main sequence, turnoff, and red giant branch), with metallicities varying from -2.0 up to 0.0, and ages from 4.0 to 15.0 Gyr. 
Their results show that both metallicity and age are important factors in the efficacy of atomic diffusion in altering stellar surface abundances. 
Of particular relevance for this study of M67 are models that were computed for solar metallicity and an age of 4.0 Gyr. Such models are shown as the solid curves in the different panels of Figure 7, which plot $\Delta$[X/H] versus stellar mass, where $\Delta$[X/H] = [X/H]$_{\rm Current}$ - [X/H]$_{\rm Initial}$ for the elements Fe, Mg, Si, and Ca.  
The filled red symbols represent the elemental abundances derived for the M67 stars and the pristine Fe-abundance for M67 is taken here to be the mean of the K-giants, with this value then used as the fiducial point (i.e., $\delta$[Fe/H] = 0.00) for the initial cluster value. It should be noted, however, that the abundances of red-clump stars are increased due to the reduction of the H-budget caused by H-burning but this effect is estimated as being very small (a difference of less than 0.01 dex; \citealt{Dotter2017}) and that a better initial cluster abundance would be achieved using M-dwarf abundances. (The M dwarfs in M67 would be too faint ($H$ $\sim$ 15) to be easily observed with APOGEE).

As can be seen from Figure 7, the values of $\Delta$[Fe/H] as a function of stellar mass predicted by the \cite{Dotter2017} models (with [Fe/H] = 0.0 and age = 4.0 Gyr) exhibit a behavior that is very similar to that found for the M67 stars as a function of their estimated masses. 
The behavior of the other elements Ca, Mg and Si are reminiscent of Fe: there is a pronounced abundance variation as a function of mass in overall agreement with the atomic diffusion models. However, for Mg and Si the abundances for the red-clump stars are $\sim$ 0.10 dex higher than the expectations of the models. In addition, the amplitude of the MSTO dip is largest for Ca (given the relatively high Ca abundance in the solar-type stars, which is not in good agreement with the models), while for Mg and Si the MSTO dip is less pronounced and in better agreement with the models.

Elemental abundances derived in the M67 stars covering a range of evolutionary phases suggest that diffusion processes are at work and have been observed in this cluster. Although, the number of stars in this initial APOGEE boutique sample is small, the comparison of spectroscopically derived Fe abundances with those predicted by stellar models that include atomic diffusion is very promising and this pilot study demonstrates what can be accomplished with the APOGEE spectra. 
Souto et al. (\textit{in preparation}) will present abundances for a much larger sample of M67 members based on the stellar parameters obtained from photometry and isochrones and chemical abundances derived automatically from the APOGEE spectra; these results will also be compared with the DR14 abundances.

\begin{figure*}
\figurenum{7}
\begin{center}
\includegraphics[width=1\linewidth]{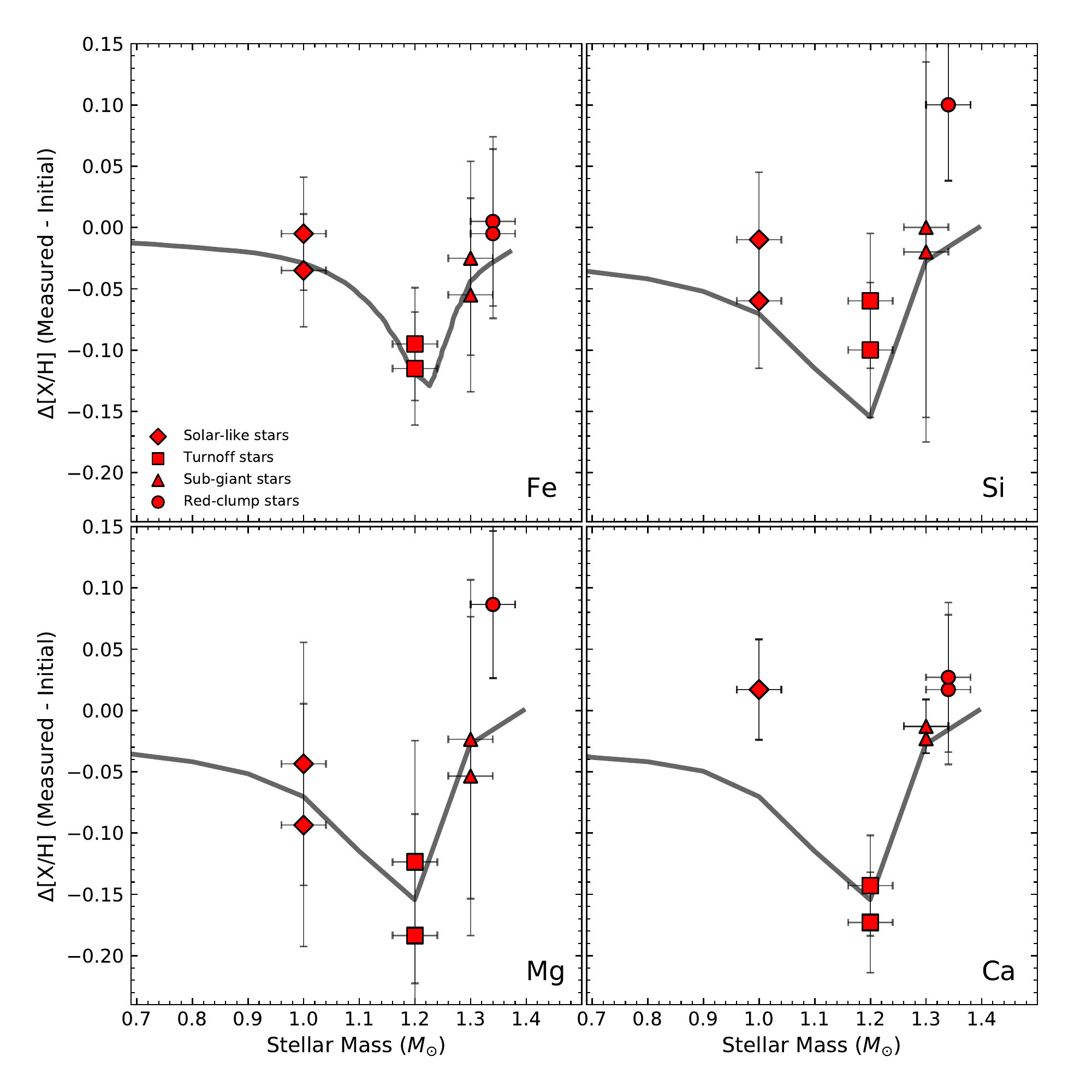}
\caption{Atomic diffusion models (solar metallicity and 4.0 Gyr; \citealt{Dotter2017}, \citealt{Choi2016}) for the stellar mass as a function of $\Delta$[X/H] are shown as the grey lines, where $\Delta$[X/H] indicates the current stellar photospheric abundances of the elements (Fe, Mg, Si, and Ca) minus the initial cluster composition. The filled red symbols represent the elemental abundances derived for the M67 stars.
}
\end{center}
\label{fig7:fig7}
\end{figure*}

\section{Summary}

This paper presents detailed chemical abundances for the elements C, N, O, Na, Mg, Al, Si, K, Ca, Ti, V, Cr, Mn, and Fe in eight stellar members of the open cluster M67. The sample stars have different stellar masses and are in different evolutionary stages: two G-dwarfs, two turnoff stars, two G-subgiants, and two K-giants (RC). 
Abundances were derived via a manual ``boutique'' detailed spectroscopic line-by-line abundance analysis using APOGEE high-resolution NIR spectra.

The derived abundances were investigated as a function of stellar evolutionary state, with homogeneous abundances found for each pair of stars in the same phase of evolution.  
Significant changes in abundance ($\sim$ 0.05 - 0.20 dex) were found across different evolutionary states, with most of the studied elements (except K, Ti and Cr) having their lowest values in the turnoff stars, while the red clump stars tended to exhibit the largest abundances. 
For iron, in particular, we obtain  $\langle$A(Fe)$\rangle$ = 7.48 $\pm$ 0.05 for the solar-like stars, $\langle$A(Fe)$\rangle$ = 7.39 $\pm$ 0.05 for turnoff stars, $\langle$A(Fe)$\rangle$ = 7.46 $\pm$ 0.08 for subgiants stars, and $\langle$A(Fe)$\rangle$ = 7.50 $\pm$ 0.07 for the red clump stars. 

We have conducted a comparison of the derived abundances of Mg, Si, Ca, and Fe as a function of stellar mass with atomic diffusion models in the literature (\citealt{Dotter2016, Dotter2017}, \citealt{Choi2016}). 
We find that the atomic diffusion models explain reasonably well most of the observed abundance trends suggesting that the signature of diffusion has been detected in M67 stars.

Published departures from LTE-derived abundances from Na I, Mg I, and Si I lines in the APOGEE spectral region were examined and are expected to be significantly smaller than the changes in abundance found in the M67 stars: non-LTE is unlikely to explain the observed abundance spreads that correlate with evolutionary state, but 3-D calculations are still needed.
Diffusion effects will be investigated further using the entire set of APOGEE spectra and ASPCAP results for M67 members in a follow-up paper.

\acknowledgments

We thank the anonymous referee for useful comments that helped improve the paper.
We thank Jo Bovy and Dennis Stello for useful comments. 
KC thanks Corinne Charbonnel for fruitful conversations and Karin Lind for checking non-LTE corrections for Na.
KC and VS acknowledge that their work here is supported, in part, by the National Aeronautics and Space Administration under Grant 16-XRP16\_2-0004, issued through the Astrophysics Division of the Science Mission Directorate. D.A.G.H. was funded by the Ram\'on y Cajal fellowship number  RYC-2013-14182. D.A.G.H. and O.Z. acknowledge support  provided  by  the  Spanish  Ministry  of  Economy  and  Competitiveness (MINECO) under grant AYA-2014-58082-P. P.M.F. acknowledges support provided by the National Science Foundation (NSF) under grants AST-1311835 and AST-1715662. SV gratefully acknowledges the support provided by Fondecyt reg. n. 1170518. H. J\"onsson acknowledges support from the Crafoord Foundation, and Stiftelsen Olle Engkvist Byggm\"astare.
Funding for the Sloan Digital Sky Survey IV has been provided by the Alfred P. Sloan Foundation, the U.S. Department of Energy Office of Science, and the Participating Institutions. SDSS-IV acknowledges
support and resources from the Center for High-Performance Computing at the University of Utah. The SDSS web site is www.sdss.org.

SDSS-IV is managed by the Astrophysical Research consortium for the 
Participating Institutions of the SDSS Collaboration including the 
Brazilian Participation Group, the Carnegie Institution for Science, 
Carnegie Mellon University, the Chilean Participation Group, the French Participation Group, Harvard-Smithsonian Center for Astrophysics, 
Instituto de Astrof\'isica de Canarias, The Johns Hopkins University, 
Kavli Institute for the Physics and Mathematics of the Universe (IPMU) / 
University of Tokyo, Lawrence Berkeley National Laboratory, 
Leibniz Institut f\"ur Astrophysik Potsdam (AIP),  
Max-Planck-Institut f\"ur Astronomie (MPIA Heidelberg), 
Max-Planck-Institut f\"ur Astrophysik (MPA Garching), 
Max-Planck-Institut f\"ur Extraterrestrische Physik (MPE), 
National Astronomical Observatory of China, New Mexico State University, 
New York University, University of Notre Dame, 
Observat\'orio Nacional / MCTI, The Ohio State University, 
Pennsylvania State University, Shanghai Astronomical Observatory, 
United Kingdom Participation Group,
Universidad Nacional Aut\'onoma de M\'exico, University of Arizona, 
University of Colorado Boulder, University of Oxford, University of Portsmouth, 
University of Utah, University of Virginia, University of Washington, University of Wisconsin, 
Vanderbilt University, and Yale University.

\facilities{Sloan}
\software{Turbospectrum \citep{Plez2012}, \citep{AlvarezPLez1998}, Numpy \citep{numpy}, Matplotlib \citep{matplotlib}}

\bibliographystyle{yahapj}
\bibliography{ms}

\startlongtable
\begin{deluxetable}{lcccccccccccc}
\tabletypesize{\tiny}
\tablecaption{Individual Abundances}
\tablewidth{0pt}
\tablehead{
\colhead{Element} &
\colhead{$\lambda$ (\AA{})} &
\colhead{2M08510076} &
\colhead{2M08512314} &
\colhead{2M08514122} &
\colhead{2M08505182} &
\colhead{2M08513540} &
\colhead{2M08514474} &
\colhead{2M08521856} &
\colhead{2M08514388} & 
\colhead{Sun} \\
\colhead{} &
\colhead{} &
\colhead{+1153115} &
\colhead{+1154049} &
\colhead{+1154290} &
\colhead{+1156559} &
\colhead{+1157564} &
\colhead{+1146460} &
\colhead{+1144263} &
\colhead{+1156425} &
\colhead{This Work} \\
\colhead{} &
\colhead{} &
\colhead{G-dwarf} &
\colhead{G-dwarf} &
\colhead{G-turn-off} &
\colhead{G-turn-off} &
\colhead{G-subgiant} &
\colhead{G-subgiant} &
\colhead{K-giant} &
\colhead{K-giant} &
\colhead{} &
\colhead{} 
}
\startdata
{\bf Fe I}	
&15194.492	&...	&...	&...	&...	&7.50	&7.45	&7.55	&7.53	&...\\
&15207.528	&7.54	&7.49	&7.36	&7.39	&7.47	&7.46	&7.51	&7.48	&7.45\\
&15220	 	&7.44	&7.51	&7.32	&7.37	&7.46	&7.43	&...	&...	&7.45\\ 
&15224	 	&7.53	&7.62	&7.40	&7.42	&7.50	&7.45	&...	&...	&7.52\\
&15240	 	&7.51	&7.53	&7.43	&7.43	&7.47	&7.44	&...	&...	&7.49\\ 
&15245	 	&7.48	&7.54	&7.41	&7.42	&7.47	&7.46	&...	&...	&7.46\\ 
&15294		&7.42	&7.43	&7.33	&7.30	&7.47	&7.42	&...	&...	&7.43\\
&15301	 	&7.50	&7.47	&7.42	&7.43	&7.52	&7.48	&...	&...	&7.47\\ 
&15344		&7.47	&7.49	&7.43	&7.42	&7.52	&7.48	&...	&...	&7.51\\ 
&15395.718	&7.50	&7.49	&7.42	&7.42	&7.48	&7.48	&7.54	&7.52	&7.47\\ 
&15490.339	&7.53	&7.54	&...	&7.45	&7.45	&7.48	&7.52	&7.48	&7.46\\
&15498	 	&7.51	&7.50	&7.45	&7.43	&7.49	&7.48	&...	&...	&7.43\\ 
&15502	 	&7.47	&7.53	&7.44	&7.34	&7.50	&...	&...	&...	&7.41\\ 
&15532		&7.49	&7.52	&7.34	&7.37	&7.46	&7.42	&...	&...	&7.49\\ 
&15534	 	&...	&...	&...	&...	&...	&...	&...	&...	&7.44\\ 
&15537	 	&...	&...	&7.34	&7.35	&7.56	&7.44	&...	&...	&7.46\\ 
&15588		&7.41	&7.49	&7.38	&7.36	&7.48	&7.41	&...	&...	&7.43\\ 
&15648.515	&...	&...	&...	&...	&...	&...	&7.50	&7.53	&7.46\\
&15662	 	&7.37	&7.41	&7.37	&7.37	&7.46	&7.40	&...	&...	&7.41\\ 
&15677	 	&7.45	&7.47	&7.39	&7.35	&7.47	&7.45	&...	&...	&7.47\\ 
&15685	 	&7.47	&7.49	&7.36	&7.34	&7.44	&7.45	&...	&...	&7.45\\ 
&15692.751	&7.42	&7.50	&7.44	&7.44	&7.44	&7.47	&...	&...	&7.42\\
&15904	 	&7.38	&7.52	&7.36	&7.33	&7.43	&7.39	&...	&...	&7.46\\ 
&15908	 	&7.40	&7.45	&7.36	&7.34	&7.41	&7.38	&...	&...	&7.45\\ 
&15910	 	&7.37	&7.42	&7.37	&7.35	&7.46	&7.49	&...	&...	&7.47\\ 
&15913	 	&...	&...	&...	&...	&...	&...	&...	&...	&...\\ 
&15920	 	&7.41	&7.45	&7.36	&7.32	&7.41	&7.42	&...	&...	&7.42\\ 
&15964.867	&...	&...	&...	&...	&7.42	&7.40	&7.45	&7.44	&7.41\\
&15980		&7.43	&7.47	&7.41	&7.38	&7.40	&7.40	&...	&...	&7.42\\ 
&16006	 	&7.44	&7.41	&7.37	&7.31	&7.44	&...	&...	&...	&7.43\\ 
&16009.615	&7.47	&7.47	&7.41	&7.37	&7.50	&7.43	&...	&...	&7.48\\
&16038	 	&7.41	&7.48	&7.36	&7.38	&7.43	&7.44	&...	&...	&7.37\\ 
&16040.657	&7.47	&7.51	&7.38	&7.35	&7.48	&7.46	&7.47	&7.47	&7.45\\ 
&16043	 	&7.46	&7.51	&7.45	&7.35	&7.47	&7.45	&...	&...	&7.48\\ 
&16075	 	&...	&7.43	&7.37	&7.39	&7.51	&7.42	&...	&...	&7.47\\ 
&16088	 	&7.45	&7.48	&7.40	&7.38	&7.46	&7.45	&...	&...	&7.47\\ 
&16102	 	&7.45	&7.50	&7.37	&7.36	&7.48	&7.43	&...	&...	&7.48\\ 
&16115		&7.40	&7.48	&7.46	&7.43	&7.46	&7.46	&...	&...	&7.51\\ 
&16126	 	&7.46	&7.49	&7.40	&7.40	&7.48	&7.46	&...	&...	&7.41\\ 
&16153.247	&7.42	&...	&7.39	&7.41	&7.43	&7.44	&7.46	&7.48	&7.47\\ 
&16165.032	&...	&...	&...	&...	&...	&...	&7.51	&7.49	&...\\ 
&16175	 	&7.41	&7.49	&7.39	&7.38	&7.45	&7.41	&...	&...	&7.43\\ 
&16178	 	&7.44	&7.49	&7.37	&7.37	&7.46	&7.43	&...	&...	&7.48\\ 
&16180	 	&7.46	&7.51	&7.38	&7.37	&7.46	&7.44	&...	&...	&7.49\\ 
&16184	 	&...	&7.49	&7.37	&7.39	&7.46	&7.44	&...	&...	&7.50\\ 
&16195	 	&7.48	&7.46	&7.36	&7.40	&7.47	&7.42	&...	&...	&7.45\\ 
&16197	 	&7.53	&7.52	&7.46	&7.35	&7.44	&7.40	&...	&...	&7.52\\ 
&16199	 	&7.49	&7.49	&7.45	&7.44	&7.44	&7.42	&...	&...	&7.48\\ 
&16204	 	&7.40	&7.47	&...	&...	&7.42	&7.40	&...	&...	&7.43\\ 
&16207	 	&7.53	&7.48	&7.40	&7.36	&7.44	&7.40	&...	&...	&7.41\\ 
&16213	 	&7.46	&7.48	&7.45	&7.38	&7.48	&7.45	&...	&...	&7.45\\ 
&16232	 	&...	&7.49	&7.40	&7.37	&7.39	&7.44	&...	&...	&7.42\\
&16235	 	&7.46	&7.49	&7.38	&7.38	&7.47	&...	&...	&...	&7.42\\ 
&16246	 	&7.47	&7.47	&7.40	&7.36	&7.45	&7.47	&...	&...	&7.46\\ 
&16294	 	&7.52	&...	&7.37	&7.39	&7.44	&7.50	&...	&...	&7.47\\ 
&16315	 	&...	&7.51	&7.41	&7.36	&7.42	&7.45	&...	&...	&7.45\\ 
&16324	 	&7.43	&7.51	&7.33	&7.36	&7.48	&7.44	&...	&...	&7.43\\ 
&16332	 	&...	&...	&7.43	&7.46	&7.48	&7.45	&...	&...	&7.52\\ 
&16395	 	&7.50	&7.54	&...	&7.38	&7.48	&7.47	&...	&...	&7.48\\ 
&16398	 	&7.51	&...	&7.44	&...	&7.49	&7.47	&...	&...	&7.46\\ 
&16404	 	&7.50	&7.53	&7.43	&7.41	&7.49	&7.44	&...	&...	&7.49\\
&16487	 	&7.42	&7.41	&7.38	&7.39	&7.43	&7.45	&...	&...	&7.37\\ 
&16506	 	&7.53	&...	&7.44	&7.41	&7.52	&7.44	&...	&...	&7.47\\ 
&16516	 	&7.47	&7.53	&7.39	&7.40	&7.45	&7.43	&...	&...	&7.44\\ 
&16519	 	&7.50	&7.50	&7.40	&7.40	&7.49	&7.42	&...	&...	&7.45\\ 
&16522	 	&7.47	&7.52	&7.39	&7.39	&7.51	&7.46	&...	&...	&7.46\\ 
&16525		&7.47	&7.48	&7.38	&7.36	&7.50	&7.43	&...	&...	&7.44\\ 
&16531	 	&7.51	&7.57	&7.42	&7.38	&7.49	&7.42	&...	&...	&7.48\\ 
&16542	 	&7.45	&...	&7.45	&7.43	&7.48	&7.43	&...	&...	&7.45\\ 
&16552	 	&7.46	&7.47	&7.45	&7.38	&7.49	&7.43	&...	&...	&7.41\\ 
&16560	 	&7.48	&...	&7.45	&7.41	&7.49	&7.41	&...	&...	&7.40\\ 
&16612	 	&7.49	&7.48	&7.32	&7.40	&7.48	&7.43	&...	&...	&7.45\\ 
&16645	 	&7.52	&7.49	&7.43	&7.38	&7.49	&7.43	&...	&...	&7.47\\ 
&16653	 	&7.52	&7.51	&7.41	&7.39	&7.46	&7.45	&...	&...	&7.44\\ 
&16657	 	&7.48	&7.54	&7.42	&7.43	&7.50	&7.47	&...	&...	&7.47\\ 
&16661	 	&7.53	&...	&7.44	&7.45	&7.54	&7.49	&...	&...	&7.43\\ 
&16664	 	&7.48	&7.52	&7.41	&7.38	&7.49	&7.44	&...	&...	&7.49\\
 & & & & & & & &\\
{\bf CO}&	
15570-15600		&...	&...	&...	&...	&8.31	&8.33	&8.34	&8.39	&...	\\
&15970-16010	&...	&...	&...	&...	&...	&...	&8.35	&8.37	&...	\\	
&16184			&...	&...	&...	&...	&...	&...	&8.33	&8.35	&...	\\
&16600-16650	&...	&...	&...	&...	&...	&...	&...	&...	&...	\\
 & & & & & & & &\\
{\bf C I}
&15784.7		&8.41	&8.40	&8.26	&8.27	&8.31	&8.34	&...	&...	&8.31	\\
&16005.0		&8.40	&8.37	&8.28	&8.36	&8.42	&8.48	&...	&...	&8.37	\\
&16021.7		&8.41	&8.41	&8.31	&8.36	&8.34	&bad	&...	&...	&8.44	\\
 & & & & & & & &\\
{\bf CN}&	
 15260.			&...	&...	&...	&...	&8.15	&7.95	&8.03	&8.05	&...	\\		
&15322.			&...	&...	&...	&...	&...	&7.93	&8.16	&8.05	&...	\\
&15397.			&...	&...	&...	&...	&8.04	&8.02	&8.17	&8.07	&...	\\
&15332.			&...	&...	&...	&...	&...	&...	&8.10	&8.07	&...	\\
&15410.			&...	&...	&...	&...	&...	&7.93	&8.15	&8.09	&...	\\
&15447.			&...	&...	&...	&...	&8.07	&7.93	&8.17	&8.12	&...	\\
&15466.			&...	&...	&...	&...	&8.03	&7.95	&8.16	&8.06	&...	\\
&15472.			&...	&...	&...	&...	&...	&7.90	&8.18	&8.07	&...	\\
&15482.			&...	&...	&...	&...	&...	&7.96	&8.17	&8.06	&...	\\
&15580.88		&...	&...	&...	&...	&8.09	&8.05	&8.18	&8.09	&...	\\
 & & & & & & & &\\
{\bf OH}  		
&15278.334	&...	&...	&...	&...	&8.60	&8.63	&...	&...	&...\\
&15568.780	&...	&...	&...	&...	&8.66	&8.68	&...	&...	&...\\
&16190.263	&...	&...	&...	&...	&8.68	&8.65	&...	&...	&...\\
&16192.208	&...	&...	&...	&...	&8.63	&8.65	&...	&...	&...\\
 & & & & & & & &\\
{\bf Na I}		
&16373.853	&...	&...	&...	&...	&6.37	&6.40	&6.40	&6.37	&...	\\
&16388.858	&6.33	&...	&6.29	&6.36	&6.38	&6.42	&6.41	&6.46	&6.31	\\
 & & & & & & & &\\
{\bf Mg I}
&15740.716	&7.43	&7.48	&7.45	&7.33	&7.49	&7.45	&7.58	&7.60	&7.45\\
&15748.988	&7.44	&7.48	&7.38	&7.34	&7.50	&7.48	&7.65	&7.62	&7.44\\
&15765.842	&7.41	&7.46	&7.36	&7.35	&7.52	&7.48	&7.54	&7.52	&7.46\\
&15879.5	&...	&...	&...	&...	&...	&...	&7.50	&7.57	&...\\
&15886.2	&...	&...	&...	&...	&...	&...	&7.72	&7.71	&...\\
&15954.477	&...	&...	&...	&...	&...	&...	&7.68	&7.62	&...\\
 & & & & & & & &\\
{\bf Al I}		
&16718.957	&6.40	&6.36	&6.41	&6.37	&6.39	&6.41	&6.55	&6.56	&6.36\\
&16750.564	&6.37	&6.33	&6.32	&6.38	&6.43	&6.38	&6.57	&6.55	&6.34\\
&16763.360	&6.44	&6.47	&6.42	&6.37	&6.51	&6.46	&6.59	&...	&6.42\\
 & & & & & & & &\\
{\bf Si I}			
&15361.161	&...	&...	&...	&...	&...	&...	&7.62	&7.63	&...\\
&15376.831	&...	&...	&...	&...	&...	&...	&7.61	&7.59	&...\\
&15888.410	&7.46	&7.52	&7.46	&7.36	&7.52	&7.48	&...	&...	&7.42\\
&15960.063	&7.43	&7.45	&7.45	&7.39	&7.53	&7.52	&...	&...	&7.46\\
&16060.009	&...	&...	&...	&...	&...	&...	&7.56	&7.59	&...\\
&16094.787	&7.45	&7.44	&7.46	&7.39	&7.49	&7.48	&7.65	&7.63	&7.45\\
&16215.67	&...	&...	&...	&...	&...	&...	&7.63	&7.61	&...\\
&16680.770	&7.50	&7.54	&7.45	&7.52	&7.54	&7.49	&7.67	&7.64	&7.53\\
&16828.159	&...	&...	&...	&...	&...	&...	&7.60	&7.62	&...\\
 & & & & & & & &\\
{\bf K I}		
&15163.067	&5.08	&5.09	&5.07	&5.17	&5.12	&5.08	&5.14	&5.04	&5.06\\
&15168.376	&5.13	&5.16	&5.14	&5.08	&5.11	&5.06	&5.13	&5.05	&5.10\\
 & & & & & & & &\\
{\bf Ca I}		
&16136.823	&6.38	&6.38	&6.16	&6.15	&6.37	&6.32	&6.33	&6.33	&6.36\\
&16150.763	&6.32	&6.36	&...	&...	&6.30	&6.33	&6.37	&6.36	&6.31\\
&16155.236	&...	&...	&...	&...	&...	&...	&6.43	&6.39	&...\\
&16157.364	&6.37	&6.33	&6.26	&6.20	&6.36	&6.33	&6.39	&6.40	&6.32\\
 & & & & & & & &\\
{\bf Ti I}		
&15334.847	&...	&...	&...	&...	&...	&...	&...	&...	&...\\
&15543.756	&4.92	&...	&...	&...	&4.86	&4.87	&4.97	&4.91	&4.92\\
&15602.842	&...	&...	&...	&...	&...	&4.89	&5.04	&4.99	&...\\
&15698.979	&...	&...	&...	&...	&4.88	&4.92	&4.94	&4.88	&...\\
&15715.573	&4.88	&4.93	&4.96	&...	&4.85	&4.82	&5.01	&4.95	&...\\
&16635.161	&...	&...	&...	&...	&...	&...	&4.99	&4.95	&...\\
 & & & & & & & &\\
{\bf V I}		
&15924.0	&...	&...	&...	&...	&...	&...	&4.12	&4.06	&...\\
 & & & & & & & &\\
{\bf Cr I}		
&15680.063	&5.69	&5.67	&5.71	&5.64	&5.67	&5.62	&5.68	&5.67	&5.67\\
 & & & & & & & &\\
{\bf Mn I}		
&15159.0	&5.42	&5.36	&5.23	&5.21	&5.39	&5.37	&5.53	&5.55	&5.40\\
&15217.0	&5.37	&5.39	&5.31	&5.29	&5.41	&5.40	&5.53	&5.55	&5.40\\
&15262.0	&5.42	&5.41	&5.29	&5.27	&5.39	&5.38	&5.54	&5.51	&5.41\\
\tablewidth{0pt}	
\enddata
\end{deluxetable}

\begin{deluxetable}{lccccccccccc}
\tabletypesize{\tiny}
\tablecaption{Stellar Abundances}
\tablewidth{0pt}
\tablehead{
\colhead{} &
\colhead{2M08510076} &
\colhead{2M08512314} &
\colhead{2M08514122} &
\colhead{2M08505182} &
\colhead{2M08513540} &
\colhead{2M08514474} &
\colhead{2M08521856} &
\colhead{2M08514388} & 
\colhead{Sun} \\
\colhead{} &
\colhead{+1153115} &
\colhead{+1154049} &
\colhead{+1154290} &
\colhead{+1156559} &
\colhead{+1157564} &
\colhead{+1146460} &
\colhead{+1144263} &
\colhead{+1156425} &
\colhead{This Work}\\
\colhead{} &
\colhead{G-dwarf} &
\colhead{G-dwarf} &
\colhead{G-turnoff} &
\colhead{G-turnoff} &
\colhead{G-subgiant} &
\colhead{G-subgiant} &
\colhead{K-giant} &
\colhead{K-giant} &
\colhead{}
}
\startdata
Fe & 7.46 $\pm$ 0.04 &	7.49 $\pm$ 0.04	& 7.40 $\pm$ 0.04 &	7.38 $\pm$ 0.03	&	7.47 $\pm$ 0.03 &	7.44 $\pm$ 0.03 &	7.50 $\pm$ 0.03 &	7.49 $\pm$ 0.03 &	7.45 $\pm$ 0.03\\
C  & 8.41 $\pm$ 0.02 &	8.39 $\pm$ 0.02	& 8.28 $\pm$ 0.02 &	8.33 $\pm$ 0.04	&	8.35 $\pm$ 0.05 &	8.38 $\pm$ 0.07 &	8.34 $\pm$ 0.01 &	8.37 $\pm$ 0.02 &	8.37 $\pm$ 0.05\\
N  & ...			 & ... 			 	& ...			 & ... 				&	8.08 $\pm$ 0.04 &	7.96 $\pm$ 0.05 &	8.15 $\pm$ 0.04 &	8.07 $\pm$ 0.02 &	...			   \\
O  & ...			 & ... 			    & ...			 & ... 				&	...				&	...				& 	8.64 $\pm$ 0.03 &	8.65 $\pm$ 0.02 &	...				\\
Na & 6.33 			 & ...				& 6.29		 	 &	6.36 $\pm$ ...	&	6.38 $\pm$ 0.01 &	6.41 $\pm$ 0.01 &	6.41 $\pm$ 0.01 &	6.41 $\pm$ 0.07 &	6.31			\\
Mg & 7.43 $\pm$ 0.01 &	7.48 $\pm$ 0.01	&7.40 $\pm$ 0.04 &	7.34 $\pm$ 0.01	&	7.50 $\pm$ 0.02 &	7.47 $\pm$ 0.01 &	7.61 $\pm$ 0.07 &	7.61 $\pm$ 0.06 &	7.44 $\pm$ 0.01\\
Al & 6.40 $\pm$ 0.03 &	6.39 $\pm$ 0.06	&6.38 $\pm$ 0.04 &	6.37 $\pm$ 0.01	&	6.44 $\pm$ 0.05 &	6.42 $\pm$ 0.03 &	6.57 $\pm$ 0.02 &	6.55 $\pm$ 0.01 &	6.37 $\pm$ 0.04\\
Si & 7.46 $\pm$ 0.02 &	7.51 $\pm$ 0.03	&7.46 $\pm$ 0.01 &	7.42 $\pm$ 0.06	&	7.52 $\pm$ 0.02 &	7.50 $\pm$ 0.01 &	7.62 $\pm$ 0.03 &	7.62 $\pm$ 0.02 &	7.47 $\pm$ 0.04	\\
K  & 5.11 $\pm$ 0.03 &	5.13 $\pm$ 0.04	&5.11 $\pm$ 0.04 &	5.18 $\pm$ 0.05	&	5.12 $\pm$ 0.01 &	5.07 $\pm$ 0.01 &	5.14 $\pm$ 0.01 &	5.05 $\pm$ 0.01 &	5.08 $\pm$ 0.02\\
Ca & 6.37 $\pm$ 0.03 &	6.37 $\pm$ 0.02	&6.21 $\pm$ 0.05 &	6.18 $\pm$ 0.03	&	6.34 $\pm$ 0.03 &	6.33 $\pm$ 0.01 &	6.38 $\pm$ 0.04 &   6.37 $\pm$ 0.03 &	6.33 $\pm$ 0.03\\
Ti & 4.90 $\pm$ 0.02 &	4.93 $\pm$ ...	&4.96			 &...				&	4.86 $\pm$ 0.01 &	4.88 $\pm$ 0.04 &	4.99 $\pm$ 0.03 &	4.94 $\pm$ 0.04 &	4.92			\\
V  & ...			 & ... 				& ...			 & ... 				&	...				&	...				&	4.12			&	4.06			&	...				\\
Cr & 5.69		 	 &	5.67			& 5.71			 &	5.64			&	5.67			&	5.62			&	5.68			&	5.67			&	5.67			\\
Mn & 5.40 $\pm$ 0.02 &	5.39 $\pm$ 0.02	& 5.28 $\pm$ 0.04 &	5.26 $\pm$ 0.03	&	5.40 $\pm$ 0.01 &	5.38 $\pm$ 0.02 &	5.53 $\pm$ 0.01 & 	5.51 $\pm$ 0.03 &	5.40 $\pm$ 0.01	&\\ 
\tablewidth{0pt}	
\enddata
\end{deluxetable}

\clearpage
\startlongtable
\begin{deluxetable}{llccccc}
\tabletypesize{\tiny}
\tablewidth{0pt}
\tablecaption{Abundance Sensitivities due to Atmospheric Parameters}
\tablehead{
\colhead{} &
\colhead{} &
\colhead{$T_{\rm eff}$} &
\colhead{log $g$} &
\colhead{$\xi$} & 
\colhead{[M/H]} &
\colhead{$\sigma$} \\
\colhead{Stellar Class} &
\colhead{Element} &
\colhead{($+$100 K)} &
\colhead{($+$0.2 dex)} &
\colhead{($+$0.2 km s$^{-1}$)} &
\colhead{($+$0.2 dex)} &
\colhead{}}
\startdata
Red Giants &C  & +0.03	&	+0.02&	-0.03	&	+0.04	&	0.062\\
&N  & -0.04	&	+0.02&	+0.00	&	+0.08	&	0.092\\
&O  & +0.05	&	-0.03&	-0.06	&	+0.11	&	0.138\\
&Na & +0.03	&	-0.02&	+0.00	&	+0.02	&	0.041\\
&Mg & +0.04	&	-0.02&	+0.00	&	+0.04	&	0.060\\
&Al & +0.08	&	-0.02&	-0.04	&	+0.04	&	0.100\\
&Si & +0.03	&	-0.01&	-0.02	&	+0.05	&	0.062\\
&K  & +0.04	&	-0.04&	-0.02	&	+0.01	&	0.061\\
&Ca & +0.05	&	-0.02&	-0.02	&	+0.02	&	0.061\\
&Ti & +0.13	&	+0.00&	-0.01	&	+0.05	&	0.140\\
&V  & +0.06	&	+0.00&	-0.03	&	+0.03 	&	0.073\\
&Cr & +0.05	&	-0.02&	-0.03	&	+0.02	&	0.065\\
&Mn & +0.03	&	+0.02&	-0.01	&	+0.01	&	0.039\\
&Fe & +0.03	&	-0.02&	-0.05	&	+0.03	&	0.069\\
& & & & & &  \\
\hline
\hline
& & & & & &  \\
subgiants &C  &	+0.02	&	+0.02&	-0.02	&	+0.00	&	0.035\\
&N  &	-0.02	&	+0.05&	-0.03	&	+0.01	&	0.062	 \\
&O  &	+0.01	&	-0.03&	-0.02	&	+0.03	&	0.048\\
&Na &	+0.01	&	+0.01&	+0.00	&	+0.01	&	0.017\\
&Mg &	-0.06	&	-0.07&	-0.02	&	-0.09	&	0.130\\
&Al &	-0.05	&	-0.03&	-0.02	&	-0.05	&	0.079\\
&Si &	-0.05	&	-0.03&	-0.03	&	-0.04	&	0.155\\
&K  &	+0.02	&	+0.01&	+0.00	&	+0.00	&	0.077\\
&Ca &	+0.01	&	+0.01&	-0.01	&	+0.00	&	0.022\\
&Ti &	-0.02	&	-0.02&	-0.03	&	-0.03	&	0.017\\
&V  &	+0.03	&	+0.00&	-0.01	&	+0.00	&	0.014\\
&Cr &	+0.02	&	+0.02&	+0.00	&	+0.00	&	0.028\\
&Mn &	+0.02	&	+0.01&	-0.01	&	+0.00	&	0.024\\
&Fe &	+0.02	&	+0.07&	-0.03	&	+0.01	&	0.079\\
& & & & & &  \\
\hline
\hline
& & & & & &  \\
Turnoff &C  &	+0.00	&	+0.02&	+0.00	&	+0.01	&	0.022\\
&N  &			&		&			&			&		 \\
&O  &			&		&			&			&		 \\
&Na &	+0.01	&	+0.01&	+0.00	&	+0.01	&	0.017\\
&Mg &	+0.04	&	-0.04&	+0.02	&	-0.04	&	0.099\\
&Al &	+0.02	&	-0.03&	+0.02	&	-0.05	&	0.076\\
&Si &	+0.03	&	-0.03&	+0.01	&	-0.03	&	0.055\\
&K  &	+0.01	&	+0.01&	+0.00	&	+0.01	&	0.017\\
&Ca &	+0.02	&	+0.03&	+0.00	&	+0.03	&	0.041\\
&Ti &	+0.02	&	-0.03&	+0.02	&	-0.02	&	0.046\\
&V  &			&		 &			&			&		 \\
&Cr &	+0.00	&	+0.01&	+0.00	&	+0.01 	&	0.014\\
&Mn &	+0.02	&	+0.00&	+0.01	&	+0.01	&	0.024\\
&Fe &	+0.03	&	-0.02&	+0.02	&	-0.02	&	0.046\\
& & & & & &  \\
\hline
\hline
& & & & & &  \\
Solar-Type &C  &	+0.00	&	+0.02&	+0.00	&	+0.01	&	0.022\\
&N  &			&		&			&			&		 \\
&O  &			&		&			&			&		 \\
&Na &	+0.01	&	+0.01&	+0.00	&	+0.01	&	0.017\\
&Mg &	+0.03	&	-0.07&	+0.02	&	-0.06	&	0.099\\
&Al &	+0.02	&	-0.05&	+0.02	&	-0.05	&	0.076\\
&Si &	+0.02	&	-0.04&	+0.01	&	-0.03	&	0.055\\
&K  &	+0.01	&	+0.01&	+0.00	&	+0.01	&	0.017\\
&Ca &	+0.02	&	+0.02&	+0.00	&	+0.03	&	0.041\\
&Ti &	+0.02	&	-0.03&	+0.02	&	-0.02	&	0.046\\
&V  &			&		 &			&			&		 \\
&Cr &	+0.00	&	+0.01&	+0.00	&	+0.01 	&	0.014\\
&Mn &	+0.02	&	+0.00&	+0.01	&	+0.01	&	0.024\\
&Fe &	+0.02	&	-0.03&	+0.02	&	-0.02	&	0.046\\
\enddata
\end{deluxetable}

\end{document}